\documentclass[journal]{IEEEtran}
\IEEEoverridecommandlockouts
\usepackage{cite}
\usepackage{amsmath,amssymb,amsfonts}
\usepackage{graphicx}
\usepackage{textcomp}
\usepackage{subfig}
\usepackage{algorithm}
\usepackage{algorithmic}
\usepackage{amssymb}
\usepackage{math}
\usepackage{balance}
\usepackage{multirow}
\usepackage{booktabs}
\usepackage{enumitem}
\usepackage{url}

\newcommand{\ie}{i.e. }
\newcommand{\inv}{^{\raisebox{.2ex}{$\scriptscriptstyle-1$}}}

\usepackage[printonlyused,withpage]{acronym}
\acrodef{DP-RTF}{direct-path relative transfer function}
\acrodef{EG}{exponentiated gradient}
\acrodef{DPRTF-EG}{\ac{DP-RTF} using \ac{EG}}
\acrodef{GMM}{Gaussian mixture model}
\acrodef{CGMM}{complex Gaussian mixture model}
\acrodef{TDOA}{time difference of arrival}
\acrodefplural{TDOA}[TDOAs]{time differences of arrival}
\acrodef{DOA}{direction of arrival}
\acrodefplural{DOA}[DOAs]{directions of arrival}
\acrodef{SRP}{steered-response power}
\acrodef{STFT}{short-time Fourier transform}
\acrodef{TF}{time-frequency}
\acrodef{IPD}{interaural phase difference}
\acrodef{PRP}{pair-wise relative phase ratio}
\acrodef{CTF}{convolutive transfer function}
\acrodef{PHD}{probability hypothesis density}
\acrodef{RLS}{recursive least squares}
\acrodef{VEM}{variational expectation maximization}
\acrodef{RIR}{room impulse response}
\acrodef{PSD}{power spectral density}
\acrodef{EM}{expectation maximization}
\acrodef{REM}{recursive \ac{EM}}
\acrodef{SNR}{signal-to-noise ratio}
\acrodef{HRTF}{head-related transfer function}
\acrodef{MAE}{mean absolute error}
\acrodefplural{MAE}[MAEs]{mean absolute errors}
\acrodef{MD}{miss detection}
\acrodefplural{MD}[MDs]{missed detections}
\acrodef{FA}{false alarm}
\acrodefplural{FA}[FAs]{false alarms}
\acrodef{ID}{identity switch}
\acrodefplural{ID}[IDs]{identity switches}
\acrodef{PHAT}{phase transform}
\acrodef{ROC}{receiver operating characteristic}
\acrodef{LMS}{least mean squares}
\acrodef{RF}{real-time factor}
\acrodef{8ch}{eight-channel}
\acrodef{4ch}{four-channel}
\acrodef{Kinovis-MST}{Kinovis multiple speaker tracking}

\usepackage[final]{review}
\setcoverletter{coverletter-R2.tex}
\setrevision{2}

\def\BibTeX{{\rm B\kern-.05em{\sc i\kern-.025em b}\kern-.08em
    T\kern-.1667em\lower.7ex\hbox{E}\kern-.125emX}}

\begin{document}

\title{Online Localization
 and Tracking of Multiple Moving Speakers in
 Reverberant Environments}

\author{Xiaofei Li$^{*}$, Yutong Ban$^{*}$, Laurent Girin, Xavier Alameda-Pineda and Radu Horaud
      
 \thanks{$^*$ X. Li and Y. Ban have equally contributed to the paper.} 
   
 \thanks{X. Li, Y. Ban, X. Alameda-Pineda and R. Horaud are with Inria Grenoble Rh\^one-Alpes and with Univ. Grenoble Alpes, France. 
 }
 
 \thanks{L. Girin is with Univ. Grenoble Alpes, Grenoble-INP, GIPSA-lab and Inria. 
 }
 
 \thanks{This work was supported by the ERC Advanced Grant VHIA \#340113.}
 }



\maketitle

\begin{abstract}
We address the problem of online localization and tracking of multiple moving speakers in reverberant environments. The paper has the following contributions. We use the \ac{DP-RTF}, an inter-channel feature that encodes acoustic information robust against reverberation, and we propose an online algorithm well suited for estimating \acp{DP-RTF} associated with moving audio sources.
Another crucial ingredient of the proposed method is its ability to properly assign \acp{DP-RTF} to audio-source directions. Towards this goal, we adopt a maximum-likelihood formulation and we propose to use  \ac{EG} to efficiently 
update source-direction estimates starting from their currently available values. 
The problem of multiple speaker tracking is computationally intractable because the number of possible associations between observed source directions and physical speakers grows exponentially with time. We adopt a Bayesian framework and we propose a variational approximation of the posterior filtering distribution associated with multiple speaker tracking, as well as an efficient \ac{VEM} solver. 
The proposed online localization and tracking method is thoroughly evaluated using two datasets that contain recordings performed in real environments.
\end{abstract}


\section{Introduction}
The localization and tracking of multiple speakers in real world environments are very challenging tasks,  in particular in the presence of reverberation and ambient noise and of natural conversations, e.g. 
short sentences, speech pauses and frequent speech turns among speakers. 
Methods based on \acp{TDOA} between microphones, such as generalized cross-correlation \cite{knapp1976}, are typically used for single-speaker localization, e.g.\cite{chen2006}.
In the case of multiple speakers, beamforming-based methods, e.g.  \ac{SRP} \cite{dibiase2001}, and subspace methods, e.g. multiple signal classification (MUSIC) \cite{ishi2009}, are widely used.
The W-disjoint orthogonality (WDO) principle \cite{yilmaz2004} assumes that the audio signal is dominated by a single audio source in small regions of the \ac{TF} domain. This assumption is particularly valid in the case of speech signals. Applying the \ac{STFT}, or any other \ac{TF} representation, inter-channel localization features, such as the \acp{IPD} \cite{yilmaz2004}, can be extracted.
In \cite{yilmaz2004}, multiple-speaker localization is based on the histogram of inter-channel features, which  is  suitable only  in the case where there is no wrapping of phase measures. In \cite{mandel2010}, a \ac{GMM} is used as a generative model of the inter-channel features of multiple speakers, with each \ac{GMM} representing one speaker, and each \ac{GMM} component representing one candidate inter-channel time delay. An \ac{EM} algorithm iteratively estimates the component weights and assigns the features to their corresponding candidate time delays. This method overcomes the phase ambiguity problem by jointly considering all frequencies in the likelihood maximization procedure. After maximizing the likelihood, the azimuth of each speaker is given by the component that has the highest weight in the corresponding \ac{GMM}.
The complex-valued version of \ac{IPD}, i.e. the \ac{PRP}, is used in \cite{dorfan2015}. Instead of setting one \ac{GMM} for each speaker, a single \ac{CGMM} is used for all speakers with each component representing one candidate speaker location. After maximizing the likelihood of the \ac{PRP} features, with an \ac{EM} algorithm, the weight of each component represents the probability that there is an active speaker at the corresponding candidate location. Therefore, for an unknown number of speakers, counting and localization of active speakers can be jointly carried out by selecting components with large weights. 

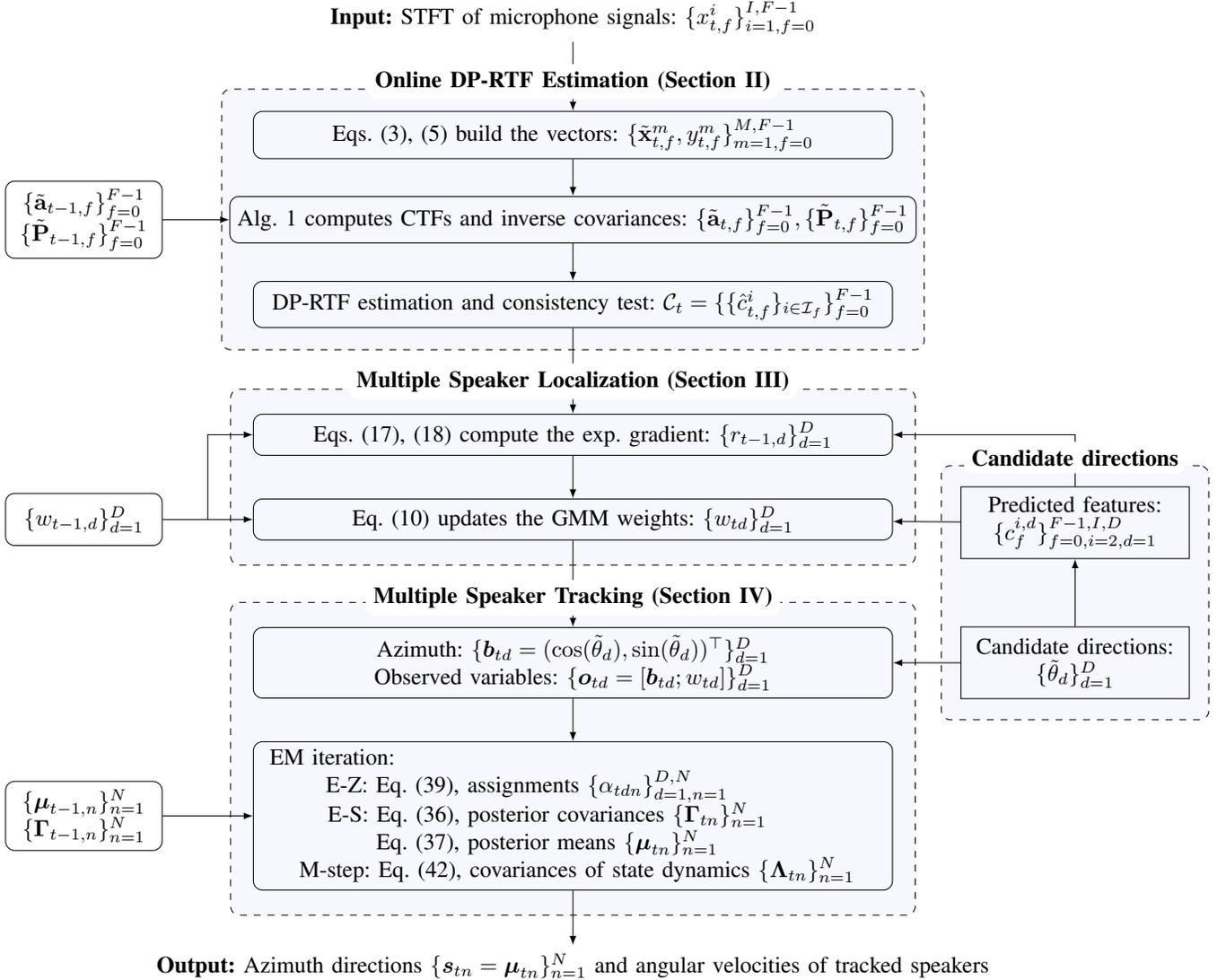
\begin{figure*}[t!]
\centering
\pgfdeclarelayer{background}

\pgfsetlayers{background}

\usetikzlibrary{shapes.geometric,backgrounds,arrows,calc,fit}
\tikzstyle{io} = [rectangle, rounded corners, minimum width=2.5cm, minimum height=1.0cm, text centered, draw=black,align=center]
\tikzstyle{data} = [rectangle, rounded corners, minimum width=9.8cm, minimum height=0.5cm, text centered, draw=black,align=center]
\tikzstyle{data1} = [rectangle, rounded corners, minimum width=2.4cm, minimum height=0.8cm, text centered, draw=black,align=center]
\tikzstyle{nobox} = [rectangle, rounded corners, draw=none,fill=white,minimum width=1cm, minimum height=0cm, text centered, align=center]
 \tikzstyle{constant} = [rectangle, minimum width=3.5cm, minimum height=1.1cm,text centered, draw=black,align=center] 
 \tikzstyle{arrow} = [->,>=latex]

\resizebox{0.99\textwidth}{!}{%

\begin{tikzpicture}


\node (xt) [nobox] {\textbf{Input:} STFT of microphone signals: $\{{x}_{t,f}^i\}_{i=1,f=0}^{I,F-1}$};

\node (t) [above of=xt,node distance=1.0cm] {};

\node (xyt) [data,below of =xt,node distance=1.8cm]{Eqs. (3), (5) build the vectors:   $\{\tilde{\mathbf{x}}^{m}_{t,f}, y^{m}_{t,f}\}_{m=1,f=0}^{M,F-1}$};

\node (at) [data,below of =xyt,node distance=1.3cm]{Alg.~1 computes CTFs and inverse covariances:   $\{\tilde{\mathbf{a}}_{t,f}\}_{f=0}^{F-1} ,\{\tilde{\mathbf{P}}_{t,f}\}_{f=0}^{F-1}$};

\node (dprtft) [data,below of =at,node distance=1.3cm]{DP-RTF estimation and consistency test: $\mathcal{C}_t=\{\{\hat{c}^i_{t,f}\}_{i\in\mathcal{I}_f}\}_{f=0}^{F-1}$};

\node (rt) [data,below of =dprtft,node distance=2.0cm]{Eqs. (17), (18) compute the exp. gradient:  $\{r_{t-1,d}\}_{d=1}^{D}$};

\node (wt) [data,below of =rt,node distance=1.3cm]{Eq. (10) updates the GMM weights: $\{w_{td}\}_{d=1}^{D}$};

\node (Ot) [data,below of =wt,node distance=2.2cm]{Azimuth: $\{\bvect_{td}=(\cos(\tilde{\theta}_{d}), \sin(\tilde{\theta}_{d}))^{\top}\}_{d=1}^{D}$ \\
Observed variables: $\{\ovect_{td} = [\bvect_{td} ; w_{td}]\}_{d=1}^{D}$};

\node (emt) [data,below of =Ot,node distance=2.35cm]{ \hspace{-7.5cm}  EM iteration:   \\ 
\hspace{-1.53cm} E-Z: Eq. (39), assignments $\{\alpha_{tdn}\}_{d=1,n=1}^{D,N}$ \\
\hspace{-0.91cm} E-S: Eq. (36), posterior covariances  $\{\Gammamat_{tn}\}_{n=1}^{N}$ \\ \hspace{-0.91cm} Eq. (37), posterior means $\{\muvect_{tn}\}_{n=1}^{N}$ \\
\hspace{-0.0cm} M-step: Eq. (42), covariances of state dynamics  $\{\Lambdamat_{tn}\}_{n=1}^{N}$}; 
\node (ot) [nobox,below of =emt,node distance=2.3cm]{\textbf{Output:} Azimuth directions $\{\svect_{tn}=\muvect_{tn}\}_{n=1}^{N}$ and angular velocities of tracked speakers};

\draw [arrow] (xt) --  (xyt);
\draw [arrow] (xyt) --  (at);
\draw [arrow] (at) --   (dprtft);
\draw [arrow] (dprtft) --  (rt);
\draw [arrow] (rt) --  (wt);
\draw [arrow] (wt) --  (Ot);
\draw [arrow] (Ot) --  (emt);
\draw [arrow] (emt) --  (ot);


\node (cand) [constant,right of =Ot,node distance=7.7cm,yshift=-0 cm]{Candidate directions: \\ $\{\tilde{\theta}_{d}\}_{d=1}^{D}$};
\node (pf) [constant,above of =cand,node distance=2.15cm]{Predicted features: \\
$\{c_f^{i,d}\}_{f=0,i=2,d=1}^{F-1,I,D}$ };

\draw [arrow] (cand) --  (pf);
\draw [arrow] (pf) |-  (rt);
\draw [arrow] (pf) --  (wt);
\draw [arrow] (cand) --  (Ot);


\node (tm1) [left of =t,node distance=2.5cm]{};
\node (vd) [below of =tm1,node distance=2.0cm]{};
\node (atm1) [data1,left of =at,node distance=7.5cm]{$\{\tilde{\mathbf{a}}_{t-1,f}\}_{f=0}^{F-1}$\\$\{\tilde{\mathbf{P}}_{t-1,f}\}_{f=0}^{F-1}$};
\node (vd) [below of =atm1,node distance=2.5cm]{};
\node (wtm1) [data1,left of =wt,node distance=7.5cm]{$\{w_{t-1,d}\}_{d=1}^{D}$};
\node (vd) [below of =wtm1,node distance=2.1cm]{};
\node (emtm1) [data1,left of =emt,node distance=7.5cm]{$\{\muvect_{t-1,n}\}_{n=1}^{N}$\\$\{\Gammamat_{t-1,n}\}_{n=1}^{N}$};
\node (vd) [below of =emtm1,node distance=2.0cm]{};

\draw [arrow] (atm1) -- node(am1) [above,align=left] { } (at);
\draw [arrow] (wtm1) -- node(wm1) [below,align=left] { } (wt) ;
\draw [arrow] (wm1) |- (rt);
\draw [arrow] (emtm1) -- node(emm1) [above] { } (emt);


\begin{scope}[on background layer]
\node[rectangle,rounded corners,draw,dashed,minimum width=10.5cm,minimum height=4.0cm] [fit = (xyt)(at)(dprtft),fill=blue!3] (bx1) {};
\end{scope}
\node (sl2) [nobox,above of =xyt,node distance=0.82cm]{\textbf{Online DP-RTF Estimation (Section II)}};

\begin{scope}[on background layer]
\node[rectangle,rounded corners,draw,dashed,minimum width=10.5cm,minimum height=2.7cm] [fit = (rt)(wt),fill=blue!3] (bx2) {};
\end{scope}
\node (sl3) [nobox,above of =rt,node distance=0.82cm]{\textbf{Multiple Speaker Localization (Section III)}};

\begin{scope}[on background layer]
\node[rectangle,rounded corners,draw,dashed,minimum width=10.5cm,minimum height=4.8cm] [fit = (Ot)(emt),fill=blue!3] (bx3) {};
\end{scope}
\node (sl3) [nobox,above of =Ot,node distance=1cm]{\textbf{Multiple Speaker Tracking (Section IV)}};

\begin{scope}[on background layer]
\node[rectangle,rounded corners,draw,dashed,minimum width=4.1cm,minimum height=3.9cm] [fit = (cand)(pf),fill=blue!3] (bx4) {};
\end{scope}
\node (cd) [nobox,above of =pf,node distance=1cm]{\textbf{Candidate directions}};

\end{tikzpicture}
}
\caption{Flowchart of the proposed multiple-speaker localization and tracking methodology.} 
\label{fig:fc}
\end{figure*}

The inter-channel features and associated localization methods mentioned above assume a direct-path propagation model: hence, they perform poorly in reverberant environments. To overcome this limitation, several \ac{TDOA} estimators based on system identification were proposed in \cite{huang2003adaptive,doclo2003,dvorkind2005,kowalczyk2013}.
 In \cite{li2017taslp} it is proposed to use the \ac{DP-RTF} as a \ac{TF}-domain inter-channel localization feature robust against reverberation. The estimation of the \ac{DP-RTF} is based on the identification of the \ac{RIR} in the \ac{STFT}-domain, i.e. the \ac{CTF}    \cite{avargel2007,talmon2009}. Overall, the method of  \cite{li2017taslp} combines the merits of  robust \ac{TDOA} estimators  \cite{huang2003adaptive,doclo2003,dvorkind2005,kowalczyk2013} and of the WDO assumption mentioned above. 

To localize moving speakers, one-stage methods such as \ac{SRP} and MUSIC can be directly used using frame-wise spatial spectrum estimators. In contrast, methods based on inter-channel features require to assign frame-wise features to speakers in an adaptive/recursive way, e.g. the smoothed histogram method of \cite{pavlidi2013}. Similar to \cite{dorfan2015}, \cite{schwartz2014} uses one \ac{CGMM} for each predefined speaker; the model is plugged into a \ac{REM} algorithm in order to update the mixture's weights.

Speaker tracking methods are generally based on Bayesian inference which combines localization with dynamic models in order to estimate the posterior probability distribution of audio-source directions, e.g. \cite{roman2008,evers2015,ban2017}.
Kalman filtering and particle filtering were used in \cite{liang2008robust} and in \cite{vermaak2001nonlinear}, respectively, for tracking a single audio source. In order to address the problem of multiple speakers, possibly with unknown and time-varying number of speakers, additional discrete latent variables are needed, i.e. observation-to-speaker assignments, as well as speaker-birth and -death processes, e.g. \cite{ba2016line}, \cite{gebru2017audio}. 
Sampling-based methods were widely used, e.g. extended particle filtering \cite{fallon2012acoustic,valin2007robust,cevher2007acoustic}, or sequential Monte Carlo implementation of the \ac{PHD} filter \cite{vo2004tracking,ma2006tracking}. However, the computational burden of sampling-based methods can be prohibitive in practice. Under some assumptions, the multiple-target tracking \ac{GMM}-\ac{PHD} filter of \cite{vo2006} has an analytical solution and is computationally efficient: it was adopted for multiple-speaker tracking in \cite{evers2015}.  

In this paper we propose a method for the simultaneous localization and tracking of multiple moving speakers (please refer to Figure~\ref{fig:fc} for a method overview). The paper has the following original contributions:
\begin{itemize}
\item Since we deal with moving speakers or, more generally, with moving audio sources, \ac{DP-RTF} features are computed using the \textit{online} \ac{CTF} estimation framework presented in  \cite{li2018sam}, based on \ac{RLS}, rather than using the \textit{batch}  \ac{CTF} estimation of \cite{li2017taslp} which assumes static audio sources. \addnote[lms]{2}{The online \ac{RLS} algorithm has a faster convergence rate than the \ac{LMS} algorithms described in \cite{huang2003adaptive,doclo2003}. This is important when dealing with multiple moving sources, where the adaptive estimator is required to quickly switch between multiple sources and to deal with moving sources.}
\item A crucial ingredient of multiple speaker localization is to properly assign acoustic features, i.e. \acp{DP-RTF}, to audio-source directions. 
We adopt the maximum-likelihood formulation of \cite{dorfan2015}. We propose to use  \ac{EG} \cite{kivinen1997} to update the source directions from their current estimated values. 
The \ac{EG}-based recursive estimator proposed below is better suited for moving sources/speakers than the batch estimator proposed in \cite{li2017taslp}.
\item The problem of multiple speaker tracking is computationally intractable because the number of possible associations between acoustic features and sources/speakers grows exponentially with time. In this paper we adopt a Bayesian variational approximation of the posterior filtering distribution which leads to an efficient \ac{VEM} algorithm. In order to deal with a varying number of speakers, we propose a birth process which allows to initialize new speakers at any time.
\end{itemize}

\addnote[diff-sam]{1}{
This paper is an extended version of  \cite{li2018sam} which has proposed an online  \ac{DP-RTF} method that has been combined with \ac{REM} to estimate the source directions. In this paper, while we keep the  \ac{DP-RTF} method of \cite{li2018sam} we propose to use \ac{EG}. The advantages of using \ac{EG} instead of \ac{REM} are described in detail in Section~\ref{sec:eg}. Moreover, the multiple speaker tracking method is completely novel. 
}

The paper is organized as follows (please refer to Figure~\ref{fig:fc}). Section~\ref{sec:dprtf} presents the online \ac{DP-RTF} estimation method. Section~\ref{sec:eg} describes the \ac{EG}-based speaker localization method and Section~\ref{sec:tracking} describes the variational approximation of the tracker and the associated \ac{VEM} algorithm. Section~\ref{sec:experiments} presents an empirical evaluation of the method based on experiments performed with real audio recordings.  Section~\ref{sec:conclusion} concludes the paper. Supplemental materials are available on our website.\footnote{\url{https://team.inria.fr/perception/research/multi-speaker-tracking/}}

\section{Recursive Multichannel \ac{DP-RTF} estimation}\label{sec:dprtf}

\subsection{Recursive Least Squares}
\label{subsec:rls}

For the sake of clarity, we first consider the noise-free single-speaker case. In the time domain $x^i(\tau)= a^i(\tau) \star s(\tau)$ is the $i$-th microphone signal, $i=1,\dots,I$,  
where $\tau$ is the time index, $s(\tau)$ is the source signal, $a^{i}(\tau)$ is the \ac{RIR} from the source to the $i$-th microphone, and $\star$ denotes the convolution. Applying the \ac{STFT} and using the \ac{CTF} approximation, for each frequency index $f=0,\dots,F-1$ we have: 
\begin{align}\label{xn}
 x^i_{t,f}=a^i_{t,f} \star s_{t,f} =\sum_{q=0}^{Q-1} a^i_{q,f}s_{t-q,f},
\end{align}
where $x^i_{t,f}$ and $s_{t,f}$ are the \ac{STFT} coefficients of the corresponding signals, and the \ac{CTF} $a^i_{t,f}$ is a sub-band representation of $a^{i}(\tau)$. Here, the convolution is executed with respect to the frame index $t$. 
The number of \ac{CTF} coefficients $Q$ is related to the reverberation time of the \ac{RIR}. 
The first \ac{CTF} coefficient $a^i_{0,f}$ mainly consists of the direct-path information, 
thence the \ac{DP-RTF} is defined as the ratio between the first \ac{CTF} coefficients of two channels:  ${a^i_{0,f}}/{a^r_{0,f}}$, where channel $r$ is the reference channel. 

Based on the cross-relation method \cite{xu1995}, using the \ac{CTF} model of one microphone pair $(i,j)$,  we have: $x^i_{t,f}\star a^j_{t,f}=x^j_{t,f}\star a^i_{t,f}$. This can be written in vector form as: 
\begin{align}
\mathbf{x}_{t,f}^{i \top}\mathbf{a}^j_f=\mathbf{x}_{t,f}^{j \ \top}\mathbf{a}^i_f,
\end{align}
with $\mathbf{a}^i_f=(a^i_{0,f},\dots,a^i_{Q-1,f})\tp$, where $\tp$ denotes matrix/vector transpose, and $\mathbf{x}^i_{t,f}=(x^i_{t,f},\dots,x^i_{t-Q+1,f})\tp$. The \ac{CTF} vector involving all channels is defined as $\mathbf{a}_f=(\mathbf{a}_f^{1 \top},\dots,\mathbf{a}_f^{I \top})\tp$.
There is a total of $I(I-1)/2$ distinct microphone pairs, indexed by $(i,j)$ with $i=1,\dots,I-1$ and $j=i+1,\dots,I$. For each pair, we construct a cross-relation equation in terms of $\mathbf{a}_f$. For this aim, we define: 
 \begin{align}\label{eq:xij}
\mathbf{x}^{ij}_{t,f} = [\underbrace{{0},\dots,{0}}_{(i-1)Q},  \mathbf{x}_{t,f}^{j \ \top}, \underbrace{{0},\dots,{0}}_{(j-i-1)Q}, -\mathbf{x}_{t,f}^{i \ \top}, \underbrace{{0},\dots,{0}}_{(I-j)Q}]\tp. 
 \end{align}  
 Then, for each pair $(i,j)$, we have:
 \begin{align}\label{eq:xija}
 \mathbf{x}_{t,f}^{ij \ \top} \mathbf{a}_f = 0.
 \end{align}
Let's assume, for simplicity, that the reference channel is $r=1$. \addnote[constraint]{1}{To avoid the trivial solution $\mathbf{a}_f=\mathbf{0}$ of (\ref{eq:xija}), we constrain the first \ac{CTF} coefficient of the reference channel to be equal to $1$}. This is done by dividing both sides of (\ref{eq:xija}) by $a^{1}_{0,f}$ and by moving the first entry of $\mathbf{x}^{ij}_{t,f}$, denoted by $-y^{ij}_{t,f}$, to the right side of (\ref{eq:xija}), which rewrites as:
\begin{align}\label{eq:tildexija}
 \tilde{\mathbf{x}}_{t,f}^{ij \ \top} \tilde{\mathbf{a}}_f = y^{ij}_{t,f},
 \end{align}
 where 
 $\tilde{\mathbf{x}}^{ij}_{t,f}$ is $\mathbf{x}^{ij}_{t,f}$ with the first entry removed, 
 and $ \tilde{\mathbf{a}}_f$ is the relative \ac{CTF} vector:
 \begin{align}\label{eq:tak}
 \tilde{\mathbf{a}}_f = \left(\frac{\tilde{\mathbf{a}}_f^{1\top}}{a^{1}_{0,f}},\frac{\mathbf{a}_f^{2\top}}{a^{1}_{0,f}},\dots,\frac{\mathbf{a}_f^{I \top}}{a^{1}_{0,f}}\right)\tp,
 \end{align}
where  $\tilde{\mathbf{a}}^1_f=(a^1_{1,f},\dots,a^1_{Q-1,f})\tp$ denotes $\mathbf{a}^1_f$ with the first entry removed. For $i=2,\dots,I$, the \ac{DP-RTF}s appear in (\ref{eq:tak}) as the first entries of $\frac{\mathbf{a}_f^{i \top}}{a^{1}_{0,f}}$. Therefore, the \ac{DP-RTF} estimation amounts to solving (\ref{eq:tildexija}).

Equation (\ref{eq:tildexija}) is defined for one microphone pair and for one frame. In batch mode, the terms $\tilde{\mathbf{x}}_{t,f}^{ij \ \top}$ and $y^{ij}_{t,f}$ of this equation can be concatenated accross microphone pairs and frames to construct a least square formulation. For online estimation, we would like to update the $\tilde{\mathbf{a}}_f$ using the current frame $t$.  
For notational convenience, let $m=1,\dots,M$ denote the index of a microphone pair, where $M=I(I-1)/2$. Then let the superscript $^{ij}$ be replaced with $^m$. The fitting error of (\ref{eq:tildexija}) is
 \begin{align}\label{eq:err}
 e^{m}_{t,f} = y^{m}_{t,f}-\tilde{\mathbf{x}}_{t,f}^{m \ \top} \tilde{\mathbf{a}}_{f}.
 \end{align}
At the current frame $t$,  for the microphone pair $m$, RLS aims to minimize the error
\begin{align}\label{eq:cost}
 J^m_{t,f} = \sum_{t'=1}^{t-1}\sum_{m'=1}^M \lambda^{t-t'}|e^{m'}_{t',f}|^2+\sum_{m'=1}^m |e^{m'}_{t,f}|^2,
 \end{align}   
which sums up the fitting error of all the microphone pairs for the past frames and the microphone pairs up to $m$ for the current frame.
The forgetting factor $\lambda \in (0,1]$ gives a lower weight to older frames, whereas all  microphone pairs have the same weight at each frame. To minimize $J^m_{t,f}$, we set its complex derivative with respect to  $\tilde{\mathbf{a}}_{f}^{\ *}$  to zero, where $^*$ denotes complex conjugate. We obtain an estimate of $\tilde{\mathbf{a}}_{f}$ at frame $t$ for microphone pair $m$ as:
 \begin{align}\label{eq:est}
 \tilde{\mathbf{a}}^m_{t,f}  =  \mathbf{R}_{t,f}^{m \ -1}&r^m_{t,f}, 
 \end{align}
 with 
  \begin{align}\label{eq:est}
 &\mathbf{R}_{t,f}^{m} = \sum_{t'=1}^{t-1}\sum_{m'=1}^{M} \lambda^{t-t'} \tilde{\mathbf{x}}_{t',f}^{m' \ *}\tilde{\mathbf{x}}_{t',f}^{m' \ \top} +\sum_{m'=1}^m \tilde{\mathbf{x}}_{t,f}^{m' \ *}\tilde{\mathbf{x}}_{t,f}^{m' \ \top}, \nonumber \\ 
 & r^m_{t,f} =
 \sum_{t'=1}^{t-1}\sum_{m'=1}^M \lambda^{t-t'}\tilde{\mathbf{x}}^{m'\ *}_{t',f}y^{m'}_{t',f}+\sum_{m'=1}^m \tilde{\mathbf{x}}^{m'\ *}_{t,f}y^{m'}_{t,f}. \nonumber
 \end{align}
 
It can be seen that the covariance matrix $\mathbf{R}^m_{t,f}$ is computed based on the rank-one modification, thence its inverse, denoted by $\mathbf{P}^{m}_{t,f}$, can be computed using the Sherman-Morrison formula, without the need of matrix inverse. 
The recursion procedure is summarized in Algorithm~\ref{alg:rls}, where $\mathbf{g}$ is the \emph{gain vector}. The current frame $t$ is initialized with the previous frame $t-1$. At the first frame, we initialize $\tilde{\mathbf{a}}^0_{1,f}$ as zero, and $\mathbf{P}^0_{1,f}$ as the identity. 
At each frame, all microphone pairs are related to the same \ac{CTF} vector that corresponds to the current speaker direction, hence all microphone pairs should be simultaneously used to estimate the \ac{CTF} vector of the current frame. In batch mode, this can be easily implemented by concatenating the microphone pairs. However, in RLS, to satisfy the rank-one modification of the covariance matrix, we need to process the microphone pairs one by one as shown in~(\ref{eq:cost}) and Algorithm~\ref{alg:rls}. At the end of the iterations over all microphone pairs, $\tilde{\mathbf{a}}^M_{t,f}$ is the ``final'' \ac{CTF} estimation for the current frame, and is used for speaker localization.
The \ac{DP-RTF} estimates, denoted as $\tilde{c}^i_{t,f}, \ i=2,\dots,I$, are obtained from $\tilde{\mathbf{a}}^M_{t,f}$. Note that implicitly we have $\tilde{c}^1_{t,f}=1$.

\begin{algorithm}[t]  \caption{\label{alg:rls} RLS at frame $t$} 
\begin{algorithmic} 
 \STATE Input: $\tilde{\mathbf{x}}^{m}_{t,f}$, $y^{m}_{t,f}$, $m=1,\dots,M$ 
 \STATE Initialization: $\tilde{\mathbf{a}}^0_{t,f}\leftarrow\tilde{\mathbf{a}}^M_{t-1,f}$, $\mathbf{P}^0_{t,f}\leftarrow\lambda^{-1}\mathbf{P}^M_{t-1,f}$ 
 \FOR{$m=1$ to $M$} 
 \STATE $e^{m}_{t,f} = y^{m}_{t,f}-\tilde{\mathbf{x}}^{m \ \top}_{t,f} \tilde{\mathbf{a}}^{m-1}_{t,f}$
 \STATE $\mathbf{g} = \mathbf{P}^{m-1}_{t,f}\tilde{\mathbf{x}}^{m \ *}_{t,f}/(1+\tilde{\mathbf{x}}^{m \ \top}_{t,f}\mathbf{P}^{m-1}_{t,f}\mathbf{x}^{m \ *}_{t,f})$ 
 \STATE $\mathbf{P}^{m}_{t,f}=\mathbf{P}^{m-1}_{t,f}-\mathbf{g}\tilde{\mathbf{x}}^{m \ \top}_{t,f}\mathbf{P}^{m-1}_{t,f}$ 
 \STATE $\tilde{\mathbf{a}}^m_{t,f}=\tilde{\mathbf{a}}^{m-1}_{t,f}+e^{m}_{t,f}\mathbf{g}$
 \ENDFOR
 \STATE Output: $\tilde{\mathbf{a}}^M_{t,f}$, $\mathbf{P}^M_{t,f}$ 
  \end{algorithmic}
\end{algorithm}

\subsection{Multiple Moving Speakers}
\label{subsec:mms}

So far, the proposed online \ac{DP-RTF} estimation method has been presented in the noise-free single-speaker case. 
The noisy multiple-speaker case was considered in \cite{li2017taslp}, but only for static speakers, i.e. batch mode, and in the two-channel case. We summarize the principles of this method and then explain in details the present online/multi-channel extension. 

\subsubsection{Estimation of the \ac{CTF} vector}

It is reasonable to assume that the  \ac{CTF} vector doesn't vary over a few consecutive frames and that only one speaker is active within a small region in the \ac{TF} domain, due to the sparse representation of speech in this domain. Consequently, the \ac{CTF} vector can be estimated over the current frame and a few past frames. An estimated \ac{CTF} value, at each \ac{TF} bin, is then assumed to be associated with only one speaker. The \ac{CTF} vector computation in the case of multiple speakers can be carried out using the \ac{RLS} algorithm, presented in Section~\ref{subsec:rls}, by adjusting the forgetting factor $\lambda$ to yield a short memory.

\addnote[ff]{2}{The forgetting factor $\lambda$ is set to $\lambda=\frac{P-1}{P+1}$, where $P$ is the number of frames being used. To efficiently estimate the \ac{CTF} vector $\tilde{\mathbf{a}}^M_{t,f}$ of length $IQ-1$, we need $\rho \times (IQ-1)$ equations, where the parameter $\rho$ should be chosen in such a way to achieve a good tradeoff between the validity of the above assumptions and robust estimation of $\tilde{\mathbf{a}}^M_{t,f}$.  To guarantee that $\rho \times (IQ-1)$ equations are available, we need $P=\frac{\rho(IQ-1)}{I(I-1)/2}\approx \rho\frac{2Q}{I-1}$ frames. One may observe that the number of frames needed to  estimate $\tilde{\mathbf{a}}^M_{t,f}$  decreases as the number of microphones increases.}   

\subsubsection{Noise reduction}
When noise is present, especially if the noise sources are temporally/spatially correlated, the \ac{CTF} estimate can be contaminated. In addition, even in a low-noise case, many TF bins are dominated by noise due to the sparsity of speech spectra. To classify the speech frames and noise frames, and to remove the noise, we use the inter-frame spectral subtraction algorithm proposed in \cite{li2015icassp,li2018sam}. 

\addnote[ss]{1}{The cross- and auto-\ac{PSD} between the convolution vector of the microphone signals, i.e. $\mathbf{x}^i_{t,f}$, and the current frame of the reference channel, i.e. ${x}^1_{t,f}$, is computed by averaging the cross- and auto-periodograms over frames.
In the present work, we use recursive averaging:  
\begin{align}
& \phivect^i_{t,f} = \beta\phivect^i_{t-1,f}+(1-\beta)\mathbf{x}^i_{t,f}{x}^{1 \ *}_{t,f}, \quad i=1,\dots,I,
\end{align}
\addnote[beta]{2}{where the smoothing parameter $\beta$ is set to achieve a good tradeoff between low noise PSD variance and fast tracking of speech variation}. The noise frames and speech frames are classified based on the minimum statistics \cite{li2015icassp} of the \ac{PSD} of ${x}^1_{t,f}$, i.e. the first entry of $\phivect^1_{t,f}$. If the frames are well classified then the noise frames only include negligible speech power, due to the sparsity and non-stationarity of speech; the speech frames include noise power similar to the noise frames, due to the stationarity of noise. Therefore, inter-frame spectral subtraction can be performed as follows: for each speech frame, the cross- and auto-\ac{PSD} of its nearest noise frame is subtracted from its cross- and auto-\ac{PSD}, then its noise-free cross- and auto-\ac{PSD} is obtained and denoted as $\hat{\phivect}^i_{t,f}$.}  

Instead of using $\mathbf{x}^i_{t,f}$, we use $\hat{\phivect}^i_{t,f}$ to construct (\ref{eq:xij}). Correspondingly, we have a new formula (\ref{eq:xija}), which is still valid, since it is equivalent to, with noise removed, taking the cross- and auto-\ac{PSD} between both sides of the initial formula (\ref{eq:xija}) and ${x}^1_{t,f}$. In the RLS process, only the speech frames (after spectral subtraction) are used, and the noise frames are skipped. A speech frame with a preceding noise frame is initialized with the latest speech frame.

\subsubsection{Consistency test} 
In practice, a \ac{DP-RTF} estimate can sometimes be unreliable. Possible reasons are that in a small frame region, (i) the \ac{CTF} is time-varying due to a fast movement of the speakers, (ii) multiple speakers are present, (iii) only noise is present due to a wrong noise-speech classification, or (iv) only reverberation is present at the end of a speech occurrence.
In \cite{li2017taslp}, a consistency test was proposed to tackle this problem: If a small frame region indeed corresponds to one active speaker, the \acp{DP-RTF} estimated using different reference channels are consistent, otherwise the \acp{DP-RTF} are biased, with inconsistent bias values. In the present work, we use the first and second channels as references, we obtain the \ac{DP-RTF} estimates $\tilde{c}^i_{t,f}$ (with $\tilde{c}^1_{t,f}=1$) and $\bar{c}^i_{t,f}$ (with $\bar{c}^2_{t,f}=1$), respectively. Then $\tilde{c}^i_{t,f}$ and $\bar{c}^i_{t,f}/\bar{c}^1_{t,f}$ are two estimates of the same \ac{DP-RTF} ${a^i_{0,f}}/{a^1_{0,f}}$.  
\addnote[sm]{1}{To measure the similarity between these two estimates, we define the vectors $\mathbf{c}_{1,t,f}^i=(1, \tilde{c}^i_{t,f})\tp$ and $\mathbf{c}_{2,t,f}^i=(1, \bar{c}^i_{t,f}/\bar{c}^1_{t,f})\tp$, where the first entries are the \acp{DP-RTF} corresponding to ${a_{0,f}^1}/{a_{0,f}^1}=1$. The similarity is the cosine of the angle between the two unit vectors:
\begin{align}\label{eq:similarity}
d_{t,f}^i = \frac{|\mathbf{c}_{1,t,f}^{i \ H}\mathbf{c}_{2,t,f}^i|}{\sqrt{\mathbf{c}_{1,t,f}^{i \ H} \mathbf{c}_{1,t,f}^i\mathbf{c}_{2,t,f}^{i \ H} \mathbf{c}_{2,t,f}^i }},
\end{align}
where $^H$ denotes conjugate transpose. If $d_{t,f}^i \in [0,1]$ is larger than a threshold (which is fixed to 0.75 in this work) then the two estimates are  consistent, otherwise they are simply ignored}. Then, the two estimates are averaged and normalized as done in \cite{li2017taslp}, resulting in a final complex-valued feature $\hat{c}^i_{t,f}$ whose module lies in the interval $[0,1]$. 

Finally, at frame $t$, we obtain a set of features  $\mathcal{C}_t=\{\{\hat{c}^i_{t,f}\}_{i\in\mathcal{I}_f}\}_{f=0}^{F-1}$, where $\mathcal{I}_f\subseteq\{2,\dots,I\}$ denotes the set of microphone indices that pass the consistency test. Note that $\mathcal{I}_f$ is empty if frame $t$ is a noise frame at frequency $f$, or if no channel passes the consistency test. Each one of these features is assumed to be associated with only one speaker.
 
\section{Localization of Multiple Moving Speakers} 
\label{sec:eg}

In this section we describe the proposed frame-wise online multiple-speaker localizer. We start by briefly presenting the underlying complex Gaussian mixture model, followed by the recursive estimation of its parameters.

\subsection{Generative Model for Multiple-Speaker Localization}
\label{subsec:eg}

In order to associate \ac{DP-RTF} features from $\mathcal{C}_t$ with  speakers and to localize each active speaker, we adopt the generative model proposed in \cite{dorfan2015}.
Let $\mathcal{D}=\{\tilde{\theta}_{1}, \dots, \tilde{\theta}_{d}, \dots, \tilde{\theta}_{D}\}$ be a set of $D$ candidate source \textit{directions}, e.g. azimuth angles.
An observed feature $\hat{c}^i_{t,f}$ (cf. Section~\ref{sec:dprtf}), when emitted by a sound source located along the direction $\tilde{\theta}_{d}$, is assumed to be drawn from a complex-Gaussian distribution with mean $c_f^{i,d}$ and variance $\sigma^2$, i.e. $\hat{c}^i_{t,f} | d \sim\mathcal{N}_c(c_f^{i,d},\sigma^2)$.
The mean $c_f^{i,d}$ is the predicted feature at frequency $f$ for channel $i$, and is precomputed based on direct-path propagation along azimuth $\tilde{\theta}_{d}$ to the microphones. The variance $\sigma^2$ is empirically set as a constant value. The marginal density of an observed feature $\hat{c}^i_{t,f}$ (taking into account all candidate directions) is a \ac{CGMM} with each component corresponding to a candidate direction: 
\begin{equation}\label{eq:CGMM}
p(\hat{c}^i_{t,f}| \mathcal{D}) = \sum_{d=1}^Dw_d\mathcal{N}_c(\hat{c}^i_{t,f};c_f^{i,d},\sigma^2),
\end{equation}
where $w_d \geq 0$ is the prior probability (component weight) of the $d$-th component, with $\sum_{d=1}^Dw_d=1$. Let us denote the vector of weights with $\wvect=(w_1, \dots, w_D)\tp$. Note that this vector is the only free parameter of the model. 
  
Assuming that the observations in $\mathcal{C}_t$ are independent, the corresponding (normalized) negative log-likelihood function, as a function of $w_d$, is given by:
\begin{align}\label{eq:loglik}
 \mathcal{L}_t = - \frac{1}{|\mathcal{C}_t|}\sum_{\hat{c}^i_{t,f}\in\mathcal{C}_t}\text{log}\Big(\sum_{d=1}^Dw_d\mathcal{N}_c(\hat{c}^i_{t,f};c_f^{i,d},\sigma^2)\Big),
\end{align}
where $|\mathcal{C}_t|$ denotes the cardinality of $\mathcal{C}_t$.  
Once $\mathcal{L}_t$ is minimized, each weight $w_d$ represents the probability that a speaker is active in the direction  $\tilde{\theta}_{d}$. Therefore, sound source localization amounts to the minimization of $\mathcal{L}_t$. In addition, taking into account the fact that the number of actual active speakers is much lower than the number of candidate directions, an entropy term was proposed in \cite{li2017taslp} as a regularizer to impose a sparse solution for $w_d$. The entropy is defined as 
\begin{align}\label{eq:entropy}
 H = -\sum_{d=1}^Dw_d\text{log}(w_d).
\end{align}
The  concave-convex procedure \cite{yuille2003} was adopted in \cite{li2017taslp}, to minimize the objective function $\mathcal{L}+\gamma H$ w.r.t. $\wvect$, where $\mathcal{L}$ is the normalized negative log-likelihood of the \ac{DP-RTF} features of all frames, i.e. batch mode optimization, and the positive scalar $\gamma$ was used to control the tradeoff between likelihood minimization and imposing sparsity over the weights. In the batch mode, the weight vector $\wvect$ is shared across all frames. Hence this method is not suitable for moving speakers.

\subsection{Recursive Parameter Estimation}
We now describe a recursive method for updating the weight vector from $\wvect_{t-1}$ to $\wvect_{t}$, i.e. from frame $t-1$ to frame $t$, using the \ac{DP-RTF} features at $t$. This can be formulated as the following online optimization problem \cite{kivinen1997}: 
\begin{align}\label{eq:oopt}
&{\wvect}_t = \mathop{\textrm{argmin}}_{\wvect} \ \chi(\wvect,\wvect_{t-1})+ \eta(\mathcal{L}_t+\gamma H),\\
\label{eq:constraints-w}
&\text{s.t.} \quad w_d>0, \ \forall d \in \{1\dots D\} \quad \text{and} \quad \sum_{d=1}^Dw_d=1, 
\end{align}
where $\chi(\avect,\bvect)$ is a distance between $\avect$ and $\bvect$. The positive scalar factor $\eta$ controls the parameter update rate.  
To minimize \eqref{eq:oopt}, the derivative of the objective function w.r.t $\wvect$ is set to zero, yielding a set of equations with no closed-form solution.
To  speed up the computation, it is assumed that $\wvect_{t}$ is close to $\wvect_{t-1}$, thence the derivative of $\mathcal{L}_t+\gamma H$ at $\wvect$ can be approximated with the derivative of $\mathcal{L}_t+\gamma H$ at $\wvect_{t-1}$. This assumption is reasonable when parameter evolution is not too fast.  As a result, when the distance $\chi(\wvect,\wvect_{t-1})$ is Euclidean, the objective function leads to gradient descent with  a step length equal to $\eta$. Nevertheless, the constraints \eqref{eq:constraints-w} lead to an inefficient gradient descent procedure. To obtain an efficient solver, we exploit the fact that the weights $w_d$ are probability masses, hence we replace the Euclidean distance with the more suitable Kullback-Leibler divergence, \ie $\chi(\wvect,\wvect_{t-1})=\sum_{d=1}^D w_d\text{log}\frac{w_d}{w_{t-1,d}}$, which results in the exponentiated gradient algorithm \cite{kivinen1997}.

The partial derivatives of $\mathcal{L}_t$ and $H$ w.r.t $w_d$ at the point $w_{t-1,d}$ are computed with, respectively: 
\begin{align}\label{eq:grad}
\frac{\partial (\mathcal{L}_t)}{\partial w_d}\Big|_{w_{t-1,d}} &= -\frac{1}{|\mathcal{C}_t|}\sum_{\hat{c}^i_{t,f}\in\mathcal{C}_t}\frac{\mathcal{N}_c(\hat{c}^i_{t,f};c_f^{i,d},\sigma^2)}{\sum_{d'=1}^D w_{t-1,d'}\mathcal{N}_c(\hat{c}^i_{t,f};c_f^{i,d'},\sigma^2)}, \nonumber \\
\frac{\partial  H}{\partial w_d}\Big|_{w_{t-1,d}} &= -(1+\text{log}(w_{t-1,d})), \quad \forall d \in \{1\dots D\}. 
\end{align}
Then, the exponentiated gradient,
\begin{align}\label{eq:exgrad}
r_{t-1,d}=e^{-\eta\big(\frac{\partial (-\mathcal{L}_t)}{\partial w_d}\big|_{w_{t-1,d}}+\gamma\frac{\partial  H}{\partial w_d}\big|_{w_{t-1,d}}\big)}, \quad \ \forall d \in \{1\dots D\},
\end{align}
is used to update the weights with:
\begin{align}\label{eq:eg}
 w_{t,d} = \frac{r_{t-1,d}w_{t-1,d}}{\sum_{d'=1}^D r_{t-1,d'}w_{t-1,d'}}, \quad \ \forall d \in \{1\dots D\}.
\end{align}
It is clear from (\ref{eq:eg}) that the parameter constraints \eqref{eq:constraints-w} are necessarily satisfied. 
The exponentiated gradient algorithm sequentially evaluates (\ref{eq:grad}), (\ref{eq:exgrad}) and (\ref{eq:eg}) at each frame. At the first frame, the weights are initialized with the uniform distribution, namely  $w_{1,d}=\frac{1}{D}$. When $\mathcal{C}_t$ is empty, such as during a silent period, the parameters are recursively updated with $w_{t,d}=(1-\eta')w_{t-1,d}+\eta'\frac{1}{D}$. 

The weight $\wvect_{t}$ as a function of $\tilde{\theta}_{d}$, i.e.  $w_{t,d}$, exhibits a handful of peaks that could correspond to active speakers.  The use of an entropy regularization term was shown to both suppress small spurious peaks, present without using the regularization term, and to sharpen the peaks corresponding to actual active speakers, thus allowing to better localize true speakers and to eliminate erroneous ones.
In the case of moving speakers, a peak should shift along time from a direction $\tilde{\theta}_{d}$ to a nearby direction. Spatial smoothing of the weight function raises the weight values around a peak, which results in smoother peak jumps. \addnote[ss]{2}{In our experiments, spatial smoothing is carried out with $w_{t,d} = (w_{t,d}+0.02w_{t,d-1}+0.02w_{t,d+1})/1.04$, where the smoothing factor $0.02$ is empirically chosen in order to smooth peak motion from one frame to the next, while avoiding the peaks to collapse}.
One may think that spatial smoothing and entropy regularization neutralize each other, but in practice it was found that their combination is beneficial.

\subsection{Peak Selection and Frame-wise Speaker Localization}
\label{subsec:peak}
Frame-wise localization and counting of active speakers could be carried out by selecting the peaks of $\wvect_t (\tilde{\theta}_{d})$ larger than a predefined threshold \cite{li2017taslp,li2018sam}. However, peak selection does not exploit the temporal dependencies of moving speakers. Moreover, peak selection can be a risky process since a too high or too low threshold value may lead to undesirable missed detection or false alarm rates. 
In order to avoid these problems, we adopt a weighted-data Bayesian framework: all the candidate directions and the associated weights are used as observations by the multiple speaker tracking method described in Section~\ref{sec:tracking} below. The localization results obtained with peak selection are compared with the localization results obtained with the proposed tracker in Section~\ref{sec:experiments}.



\section{Multiple Speaker Tracking}

\label{sec:tracking}

%
Let $N$ be the maximum number of speakers that can  be simultaneously active at any time $t$, and let $n$ be the speaker index. Moreover, let $n=0$ denote \textit{no speaker}.  
We now introduce the main variables and their notations. Upper case letters denote random variables while lower case letters denote their realizations.

Let $\Svect_{tn}$ be a latent (or state) variable associated with speaker $n$ at frame $t$, and let $\Smat_t = (\Svect_{t1}, \dots, \Svect_{tn}, \dots, \Svect_{tN})$. 
$\Svect_{tn}$ is composed of two parts: the speaker direction and the speaker velocity. In this work, speaker direction is defined by an azimuth $\theta_{tn}$. To avoid phase (circular) ambiguity we describe the direction with the unit vector $\Uvect_{tn} = (\cos(\theta_{tn}), \sin(\theta_{tn}))^{\top}$. Moreover, let  $V_{tn}\in\mathbb{R}$ be the angular velocity. Altogether we define a realization of the state variable as $\svect_{tn} = [\uvect_{tn} ; v_{tn}]$ where the notation $[\cdot ; \cdot]$ stands for vertical vector concatenation.

Let $\Omat_t = (\Ovect_{t1}, \dots, \Ovect_{td}, \dots, \Ovect_{tD})$ be the observed variables at frame $t$. Each realization $\ovect_{td}$ of $\Ovect_{td}$ is composed of a candidate location, or azimuth $\tilde{\theta}_{td}\in\mathcal{D}$, and a weight $w_{td}$. The weight $w_{td}$ is the probability that there is an active speaker in the direction $\tilde{\theta}_{td}$, namely \eqref{eq:oopt}. As above, let the azimuth be described by a unit vector $\bvect_{td}=(\cos(\tilde{\theta}_{td}), \sin(\tilde{\theta}_{td}))^{\top}$. In summary we have $\ovect_{td} = [\bvect_{td} ; w_{td}]$.
Moreover, let  $Z_{td}$ be a (latent) assignment variable associated with each observed variable $\Ovect_{td}$, such that $Z_{td}= n$ means that the observation indexed by $d$ at frame $t$ is assigned to active speaker $n\in\{0,\dots, N\}$. Note that $Z_{td} = 0$ is a ``fake" assignment -- the corresponding observation is assigned to an audio source that is either background noise or any other source that has not yet been identified as an active speaker.

The problem at hand can now be cast into the estimation of the filtering distribution $p(\smat_t, \zvect_t | \omat_{1:t})$, and further inference of $\smat_t$ and $\zvect_t$. In this work we make two hypotheses, namely (i) that the observations at frame $t$ only depend on the assignment and state variables at $t$, and (ii) that the prior distribution of the assignment variables is independent of all the other variables. By applying the Bayes rule together with these hypotheses, and ignoring terms that do not depend on $\smat_t$ and $\zvect_t$, the filtering distribution is proportional to:
\begin{align}
p(\smat_t, \zvect_t | \omat_{1:t}) \propto  p(\omat_t|  \smat_t, \zvect_t) p(\zvect_t) p(\smat_t | \omat_{1:t-1}),
\label{eq:posterior_Bayes}
\end{align}
which contains three terms:
the observation model, the prior distribution of the assignment variable and the predictive distribution over the sources state. We now characterize each one of these three terms.

\subsubsection{Audio observation model}
\label{subsec:audio-model}
The audio observation model describes the distribution of the observations given speakers state and assignment. 
We assume the different observations are independent, conditioned on speakers state and assignment, which can be written as:
\begin{equation}
\label{eq:audio-independence}
p(\omat_t| \zvect_t, \smat_t ) = \prod_{d=1}^{D} p (\omat_{td} | \zvect_t, \smat_t).
\end{equation}
Since the weights describe the confidence associated with each observed azimuth, we adopt the weighted-data GMM model of \cite{gebru2016algorithms}:
\begin{align}
\label{eq:assign}
p (\bvect_{td} | & Z_{td}=n,  \smat_{tn}; w_{td})= \nonumber \\
&
\begin{cases}
\mathcal{N} (\bvect_{td};  \Mmat \svect_{tn} , \frac{1}{w_{td}}\Sigmamat) & \textrm{if} \:  n \in \{1,\dots,N\} \\
\mathcal{U} (\textrm{vol} (\mathcal{G})) & \textrm{if} \: n=0
\end{cases},
\end{align}
where the matrix $\Mmat =\left[ \Imat_{2 \times 2}, \textbf{0}_{2 \times 1} \right]$ projects the state variable onto the space of source directions and $\Sigmamat$ is a covariance matrix (set empirically to a fixed value in the present study). Note that the weight plays the role of a precision: The higher the weight $w_{td}$, the more reliable the source direction $\bvect_{td}$.  
The case $Z_{td} =0$ follows a uniform distribution over the volume of the observation space. 

\subsubsection{Prior distribution}
The prior distribution of the assignment variable is independent over observations and is assumed to be uniformly distributed over all the speakers (including  the case $n=0$), hence: 
\begin{equation}
\label{eq:prior}
p(\zvect_{t}) = \prod_{d = 1}^{D} p(Z_{td}=n) \quad \text{with} \quad  \pi_{dn}= p(Z_{td}=n)= \frac{1}{N+1}.
\end{equation}

\subsubsection{Predictive distribution}
The predictive distribution describes the relationship between the state $\smat_{t}$ and the past observations up to frame $t$, $\omat_{1:t-1}$.
To calculate this distribution, we first marginalize $p(\smat_t | \omat_{1:t-1})$ over $\smat_{t-1}$, writing:
\begin{align}
\label{eq:predictive distribution}
p(\smat_t | \omat_{1:t-1}) &= \int p(\smat_t | \smat_{t-1}) p( \smat_{t-1} | \omat_{1:t-1}) d\smat_{t-1},
\end{align}
where the two terms under the integral are the state dynamics and the marginal filtering distribution of the state variable at frame $t-1$, respectively. We model the state dynamics as a linear-Gaussian first-order Markov process, independent over the speakers, \ie:   
\begin{equation}
\label{eq:state_dynamic}
p(\smat_t | \smat_{t-1}) = \prod_{n=1}^{N} \mathcal{N}
(\svect_{tn};\Dmat_{t-1,n} \svect_{t-1,n},\Lambdamat_{tn}),
\end{equation}
where $\Lambda_{tn}$ is the dynamics' covariance matrix and $\Dmat_{t-1,n}$ is the state transition matrix.
Given the estimated azimuth $\theta_{t-1,n}$ and angular velocity $v_{t-1,n}$ at frame $t-1$, we have the following relation:
\begin{align}
\label{eq:dynamic relation}
\begin{pmatrix}
\cos(\theta_{tn})\\
\sin(\theta_{tn})\\
\end{pmatrix}
=
\begin{pmatrix} \cos(\theta_{t-1,n} + v_{t-1,n} \Delta t)\\
\sin(\theta_{t-1,n} + v_{t-1,n} \Delta t)
\end{pmatrix},
\end{align}
where $\Delta t$ is the time increment between two consecutive frames. Expanding \eqref{eq:dynamic relation} and assuming that the angular displacement $v_{t-1,n} \Delta t$ is small, the state transition matrix can be written as:
\begin{equation}
\label{eq:transition-matrix}
\Dmat_{t-1,n} = \begin{pmatrix}
1 & 0 &-\sin(\theta_{t-1,n}) \Delta t \\
0 & 1 & \cos(\theta_{t-1,n}) \Delta t \\
0 & 0 & 1 \\
\end{pmatrix}.
\end{equation}
In the following $\Dmat_{t-1,n}$ is written as $\Dmat$, only to lighten the equations. 

\subsection{Variational Expectation Maximization Algorithm}
At this point, the standard solution to the calculation of the filtering distribution consists of using EM methodology. EM alternates between evaluating the expected complete-data log-likelihood and maximizing this expectation with respect to the model parameters. More precisely, the expectation writes:
\begin{equation}
\label{eq:q-func}
J(\Thetavect,\Thetavect^{o}) = \mathbf{E}_{p(\zvect_t,\smat_t|\omat_{1:t},\Thetavect^{o})} \left[\log p(\zvect_t,\smat_t,\omat_{1:t}|\Thetavect) \right],
\end{equation}
where $\Thetavect^{o}$ denotes the current parameter estimates and $\Thetavect$ denotes the new estimates, obtained via maximization of  \eqref{eq:q-func}.
However, given the hybrid combinatorial-and-continuous nature of the latent space, such solution is intractable in practice, due to combinatorial explosion.
We thus propose to use of a variational approximation to solve the problem efficiently. We inspire from~\cite{ba2016line} and propose the following factorization:
\begin{equation}
\label{eq:variational-approximation}
p(\zvect_t,\smat_t|\omat_{1:t}) \approx 
q(\zvect_t,\smat_t) = q(\zvect_{t}) 
\prod_{n=0}^{N} q(\svect_{tn}).
\end{equation} 
The optimal solution is then given by two E-steps, an E-S step for each individual state variable $\Svect_{tn}$ and an E-Z step for the assignment variable $\Zvect_{t}$:
\begin{align}
\label{eq:variational-state}
\log q(\svect_{tn}) & = \Emat_{q(\zvect_{t}) \prod_{m \neq n} 
q(\svect_{tm})} [\log p(\zvect_{t}, \smat_{t} | \omat_{1:t} )],\\
\label{eq:variational-assignment}
\log q(\zvect_{t}) & = \Emat_{q(\smat_t)}  [\log p(\zvect_{t}, \smat_{t} | \omat_{1:t}) ].
\end{align}
It is easy to see that in order to compute~\eqref{eq:variational-state} and~\eqref{eq:variational-assignment}, two elements are needed: the predictive 
distribution~\eqref{eq:predictive distribution} and the marginal filtering distribution at $t-1$,
$p(\smat_{t-1}|\omat_{1:t-1})$. Remarkably, as a consequence of the factorization~\eqref{eq:variational-approximation}, we can 
replace $p(\smat_{t-1}|\omat_{1:t-1})$ with ${q}(\smat_{t-1})=\prod_{n=1}^N{q}(\svect_{t-1,n})$ in \eqref{eq:predictive distribution} and compute the predictive distribution as follows:
\begin{align}
\label{eqn:predictive-motion}
 p(\smat_{t}|\omat_{1:t-1}) & \approx \int    p(\smat_{t}|\smat_{t-1})\prod_{n=1}^N{q}(\svect_{t-1,n})d\smat_{t-1}. 
\end{align}
This predictive distribution factorizes across speakers. Moreover, one prominent feature of the proposed variational approximation is that, if the posterior distribution at time $t-1$ ${q}(\svect_{t-1,n})$ is assumed to be a Gaussian, say
\begin{equation}
\label{eq:var-posterior-previous}
{q}(\svect_{t-1,n}) = \mathcal{N}(\svect_{t-1,n}; \muvect_{t-1,n},\Gammamat_{t-1,n}),
\end{equation}
then (the approximation of) the predictive distribution \eqref{eqn:predictive-motion} is a Gaussian. More specifically, the derivation of \eqref{eqn:predictive-motion}
 leads to:
\begin{align}\label{eqn:predictive-motion-mod}
p(\smat_{tn}|\omat_{1:t-1}) =  \mathcal{N}(\smat_{tn};\Dmat\muvect_{t-1,n},\Dmat\Gammamat_{t-1,n}\Dmat^{\top}+\Lambdamat_{tn}).
\end{align}
Moreover, as we will see in the E-S-step below, the posterior distribution at time $t$, ${q}(\svect_{tn})$, is also a Gaussian.

\subsubsection{E-S step}
The computation of the variational posterior distribution $q(\svect_{tn})$, for all currently tracked speakers, is carried out by developing \eqref{eq:variational-state} as follows. We first exploit \eqref{eq:posterior_Bayes}, \eqref{eq:audio-independence}, \eqref{eq:prior} and \eqref{eqn:predictive-motion-mod} to rewrite $\log p(\zvect_{t}, \smat_{t} | \omat_{1:t} )$ in \eqref{eq:variational-state} as a sum of individual log-probabilities. Then we eliminate all terms not depending on $\svect_{tn}$ and we evaluate the expectation of the remaining terms. Because the terms not depending on $\svect_{tn}$ were disregarded, the expectation is computed only with respect to $q(\zvect_{t})$. This nicely makes the computation of $q(\svect_{tn})$ independent of the structure of $q(\svect_{tm})$ for $m \neq n$. Eventually, this yields a Gaussian distribution:
%
\begin{equation}
\label{eq:var-posterior-t}
q(\svect_{t n}) = \mathcal{N}(\svect_{t n}; \muvect_{tn},\Gammamat_{t n}),
\end{equation}
with the following parameters:
\begin{align}
\label{eq:posteriordistribution-of-Xtn-cov}
\Gammamat_{tn} & =  \ \Big(\Big(\sum_{d=1}^{D}\alpha_{tdn} w_{td}\Big) \Mmat^{\top}{\Sigmamat}\inv\Mmat  \nonumber \\
& + \Big(\Lambdamat_{tn}+\Dmat\Gammamat_{t-1,n}\Dmat^{\top}\Big)\inv \Big)\inv,  \\
\label{eq:posterior-distribution-of-Xtn-mean}
\muvect_{tn}  & =  \Gammamat_{tn} \Big( \Mmat^{\top}{\Sigmamat}\inv \Big(\sum_{d=1}^{D}\alpha_{tdn} w_{td} \bvect_{td}\Big)  \nonumber\\ 
    &+ \Big(\Lambdamat_{tn}+\Dmat\Gammamat_{t-1,n}\Dmat^{\top}\Big)\inv\Dmat\muvect_{t-1,n}
\Big),
\end{align}
where $\alpha_{tdn}= q(Z_{td} = n)$ is the variational posterior distribution of the assignment variable, which will be detailed in Section~\ref{sec:E-Z step}.  Importantly, the first two entries of $\muvect_{tn}$ in \eqref{eq:posterior-distribution-of-Xtn-mean} represent the estimated azimuth of speaker $n$. The ``final'' azimuth estimate at frame $t$ is thus given by this subvector at the end of the VEM iterations. Since we use a unit-vector representation, we normalize this vector at each iteration of the algorithm.
Finally, note that since we have shown that $q(\svect_{t-1,n})$ being Gaussian leads to $q(\svect_{t n})$ being Gaussian as well, it is sufficient to assume that $q(\svect_{1 n})$ is Gaussian, namely at $t=1$: $q(\svect_{1 n}) = \mathcal{N}(\svect_{1n}; \muvect_{1 n},\Gammamat_{1 n})$.

\subsubsection{E-Z step}
\label{sec:E-Z step}
Developing \eqref{eq:variational-assignment} with the same principles as above, one can easily find that the variational posterior distribution of the assignment variable factorizes as:
%
%
\begin{equation}
 q(\zvect_t) = \prod_{d=1}^{D} q(z_{td}).
\end{equation}
In addition, we obtain a closed-form expression for $q(z_{td})$:
\begin{equation}
\alpha_{tdn}= q(Z_{td} = n) =  \frac{\rho_{tdn} \pi_{dn}}
{\sum_{i=0}^{N}\rho_{tdi} \pi_{di}},
\label{eq:posterior distribution of audio assignment}
\end{equation}
where $\pi_{dn}$ was defined in \eqref{eq:prior}, and $\rho_{tdn}$ is given by:
\begin{align}
\rho_{tdn}  = 
\begin{cases}
\mathcal{N}(\bvect_{td}; \Mmat \muvect_{tn}, \frac{1}{w_{td}}\Sigmamat) \\
\times e^{-\frac{1}{2} \textrm{tr} 
\left( 
 w_{td} \Mmat^{\top} \Sigmamat  \inv \Mmat 
\Gammamat_{tn}\right)} 
 & \textrm{if} \: 1\leq n \leq N \\
 \mathcal{U}( \textrm{vol}(\mathcal{G})) & \textrm{if} \: n =0.
\end{cases} 
\end{align}

\subsubsection{M-step} Once the two expectation steps are executed, we maximize $J$ in \eqref{eq:q-func} with respect to the model parameters, i.e.\ the covariance matrix of the state dynamics $\Lambdamat_{tn}$. By exploiting again the proposed variational approximation, the dependency of $J$ on $\Lambdamat_{tn}$ can be written as:
%
%
\begin{align}
\nonumber J(\Lambdamat_{tn}) = \mathbf{E}_{q(\svect_{tn})}
\left[ \log \mathcal{N}(\svect_{tn}; \Dmat \muvect_{t-1,n}, \Dmat \Gammamat_{t-1,n} \Dmat^{\top} + \Lambdamat_{tn})
\right],
\end{align}
which can be further developed as:
\begin{align}
\nonumber J(\Lambdamat_{tn}) &= \log |\Dmat \Gammamat_{t-1,n} \Dmat^{\top}  + \Lambdamat_{tn}| \\
& + \textrm{Tr}\left[(\Dmat \Gammamat_{t-1,n} \Dmat^{\top}  + \Lambdamat_{tn})^{-1} \right. \times \label{eq:J-lambda}\\
&\left. ((\muvect_{tn} - \Dmat \muvect_{t-1,n}) (\muvect_{tn} - \Dmat \muvect_{t-1,n})^{\top}+ \Gammamat_{tn})\right]. \nonumber
\end{align}
By equating to zero the gradient of \eqref{eq:J-lambda} w.r.t. $\Lambdamat_{tn}$, we obtain: 
\begin{equation}
\Lambdamat_{tn} = \Gammamat_{tn} - \Dmat \Gammamat_{t-1,n} \Dmat^{\top} + (\muvect_{tn} - \Dmat \muvect_{t-1,n})
(\muvect_{tn} - \Dmat \muvect_{t-1,n})^{\top}.
\label{eq:M:lambda}
\end{equation}

\subsection{Speaker-Birth Process}
\label{sec:Speaker-Birth Process}
A birth process is used to initialize new tracks, \ie speakers that become active. We take inspiration from the birth process for visual tracking proposed in \cite{ba2016line} and adapt it to audio tracking. The general principle is the following. In a short period of time, say from $t-L$ to $t$, with $L$ being small (typically 3), we assume that at most one new (yet untracked) speaker becomes active. For each frame from $t-L$ to $t$, among the observations assigned to $n=0$ we select the one with the highest weight, and thus obtain an observation sequence $\tilde{\ovect}_{t-L:t}$. We then compute the marginal likelihood of this sequence according to our model, $\epsilon_0=p(\tilde{\ovect}_{t-L:t})$. If these observations have been generated by a speaker that has not been detected yet (hypothese $H_1$), then they are assumed to be consistent with the model, \ie exhibit smooth trajectories, and $\epsilon_0$ will be high; otherwise, \ie if they have been generated by background noise (hypothese $H_0$), they will be more randomly spread over the range of possible observations, and  $\epsilon_0$ will be low. Giving birth to a new speaker track amounts to setting a threshold $\epsilon_1$ and deciding between the two hypotheses:
\begin{equation}
\epsilon_0   {{H_1\atop >}\atop{<\atop H_0}} \epsilon_1.
\end{equation} 
This process is applied continuously over time to detect new speakers. This includes speaker track initialization at $t=1$. Note that initially all the assignment variables are set to $n=0$ (background noise), namely $Z_{1d}=0, \forall d$.

As for the computaiotn of $p(\tilde{\ovect}_{t-L:t})$, we first rewrite it as the marginalization of the joint probability of the selected observations and the state trajectory $\hat{\svect}_{t-L:t}$ of a potential speaker:
\begin{align}
\epsilon_0=& \int p(\tilde{\ovect}_{t-L:t},\hat{\svect}_{t-L:t})d\hat{\svect}_{t-L:t},
\end{align}
which, under the proposed model, is given by:
\begin{align}
\label{eqn:birth-test}
& \epsilon_0 = \\
& \int  \Big( \prod_{i=t-L+1}^t p(\tilde{\ovect}_{i}|\hat{\svect}_{i}) p(\hat{\svect}_{i}|\hat{\svect}_{i-1}) \Big)
 p(\tilde{\ovect}_{t-L}|\hat{\svect}_{t-L}) p(\hat{\svect}_{t-L}) d\hat{\svect}_{t-L:t}. \nonumber
\end{align}
All the terms in the above equation have been defined during the derivation of our model except the marginal prior distribution of the state $p(\hat{\svect}_{t-L})$, and all these terms are Gaussian. For the track-birth process, we just want to test if the trajectory of observations from $t-L$ to $t$ is coherent, and we can define here $p(\hat{\svect}_{t-L})$ as a non-informative distribution, such as a uniform distribution. In practice we choose a Gaussian distribution with a very large covariance, to ensure a closed-form solution to \eqref{eqn:birth-test}. Due to room limitation, we do not present more details. Let us just mention that in practice we set $L=3$, which enables efficient speaker birth detection.

\begin{algorithm}[t]  \caption{\label{alg:tracking_algo}Variational EM tracking} 
\begin{algorithmic} 
 \STATE Input: audio observations $\bvect_{1:t}$\; 
 \FOR{$t = 1$ to end} 
  \STATE Gather observations at frame $t$\; 
	\FOR{$iter = 1$ to $N_{\text{iter}}$}
 \STATE E-Z-step:\\ 
   \FOR {$d \in \{1,...,D\}$}
   \FOR {$n \in \{0,...,N\}$}
   \STATE Evaluate $q(Z_{td} = n)$ with \eqref{eq:posterior distribution of audio assignment}\;

    \ENDFOR
 \ENDFOR
 \STATE E-S-step:\\ 
   \FOR {$n \in \{1,...,N\}$ }
   \STATE  Evaluate $\Gammamat_{tn}$ and $\muvect_{tn} $ with \eqref{eq:posteriordistribution-of-Xtn-cov} and \eqref{eq:posterior-distribution-of-Xtn-mean};
    \ENDFOR 
   \STATE  M-step: Evaluate $\Lambdamat_{tn}$ with \eqref{eq:M:lambda};
  
 \ENDFOR

 \STATE Speaker-Birth Process (see Section \ref{sec:Speaker-Birth Process})\;
  \STATE Detect speaker activity (see Section \ref{sec:Speaker Activity Detection})\;
   \FOR {$n \in \{1,...,N\}$}
   \IF {the speaker $n$ is detected as active}
  \STATE Output the results $\muvect_{tn}$\;
  \ENDIF
  \ENDFOR
   \ENDFOR
  \end{algorithmic}
\end{algorithm}  

\subsection{Speaker Activity Detection}
\label{sec:Speaker Activity Detection}
A very interesting feature of the proposed model is that, once speaker tracks have been estimated, the posterior distribution of the assignment variables $\Zvect_t$ can be used for speech activity detection, \ie who are the active speakers at each frame, a task also referred to as \emph{speaker diarization} in the multi-speaker context. This can be formalized as testing for each frame $t$ and each speaker $n$ between the two following hypotheses: $H_1$: Speaker $n$ is active at frame $t$, and $H_0$: Speaker $n$ is silent at frame $t$. In the present work, this is done by computing the following \textit{weighted sum of weights}, averaged over a small number of frames $L'$ to take into account speaker activity inertia, and comparing with a threshold $\delta$, a test formally written as:  
\begin{align}
\sum_{i = t-L'+1}^{t} \sum_{d = 1}^{D} \alpha_{idn} w_{i}^{d}
{{H_1 \atop >}\atop{<\atop H_0}} \delta.
\end{align}  


Overall, the variational EM tracking algorithm is described in Algorithm \ref{alg:tracking_algo}.

\section{Experiments}\label{sec:experiments}

\begin{figure*}[t]
\centering
\subfloat[Ground truth]{\includegraphics[width=0.43\textwidth]{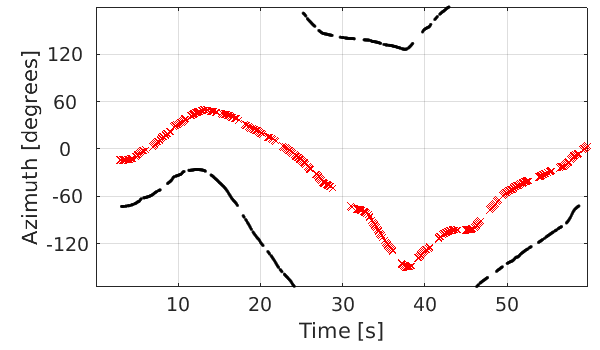}} 
\subfloat[SRP-PHAT]{\includegraphics[width=0.43\textwidth]{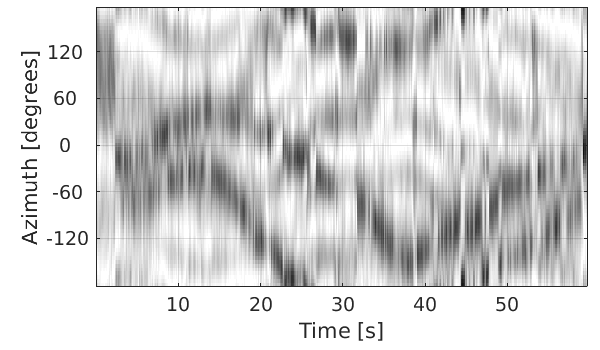}} \\
\subfloat[PRP-REM]{\includegraphics[width=0.43\textwidth]{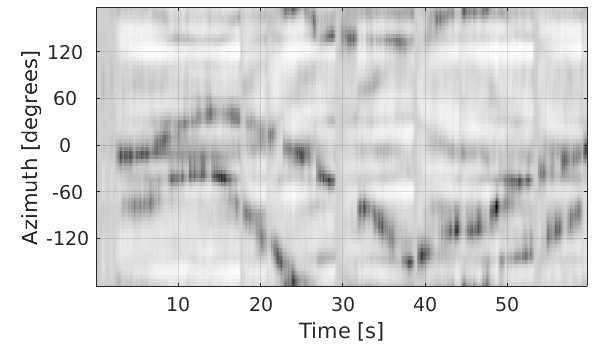}} 
\subfloat[DPRTF-REM]{\includegraphics[width=0.43\textwidth]{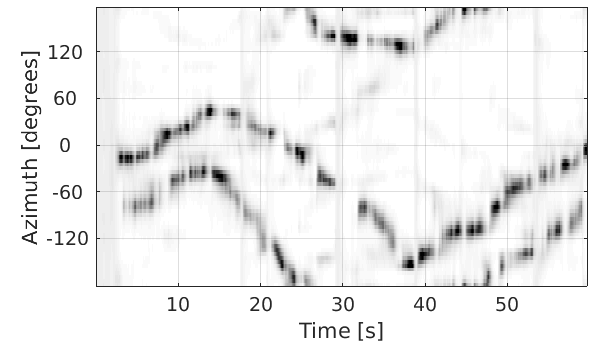}} \\
\subfloat[DPRTF-EG]{\includegraphics[width=0.43\textwidth]{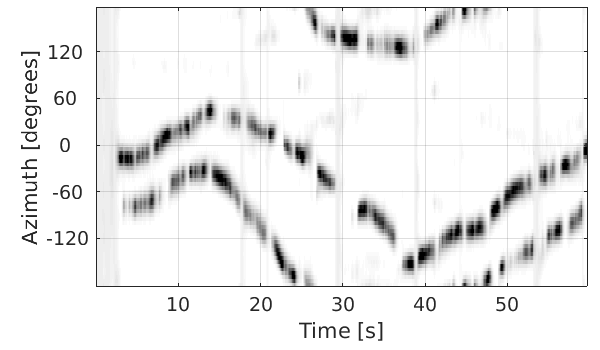}} 
\subfloat[VEM-tracking]{\includegraphics[width=0.43\textwidth]{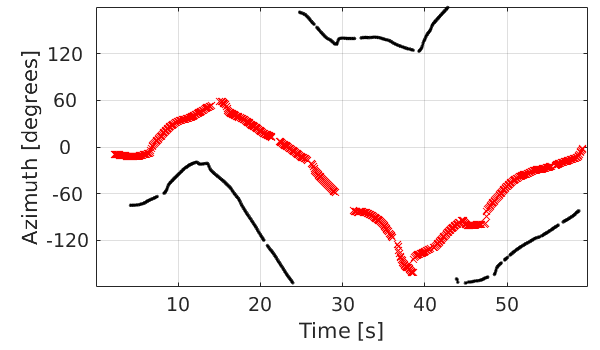}} \\
\caption{Results of speaker localization and tracking for Recording 1 / Task 6 of LOCATA data. (a) Ground truth trajectory and voice activity (red for speaker 1, black for speaker 2). Intervals in the trajectories are speaking pauses. (b)-(e) One-dimensional heat maps as a function of time for the four tested localization methods. (f) Results for the proposed \ac{VEM}-based tracker. Black and red colors demonstrate a successful tracking, \ie continuity of the tracks despite of speech pauses.}
\label{fig:locata_example}
\end{figure*}

\subsection{Experimental setups}

\subsubsection{Datasets} 
We tested and empirically validated our method with the LOCATA and the \ac{Kinovis-MST} datasets.
The LOCATA (a IEEE-AASP challenge for sound source localization and tracking) \cite{LOCATA2018a} data were recorded in the Computing Laboratory of the Department of Computer Science of Humboldt University Berlin. The room size is $7.1$~m $\times$ $9.8$~m $\times$ $3$~m, with a reverberation time $T_{60}\approx 0.55$~s. We report the results of the development corpus for tasks \#3 and \#5 with a single moving speaker, and for tasks \#4 and \#6 with two moving speakers, each task comprising three recorded sequences.\footnote{The results obtained with the proposed method were officially submitted to the LOCATA challenge and they will be available soon at \url{https://locata.lms.tf.fau.de/}.}
There are twelve microphones arranged such as to form a spherical array and placed on the head of a NAO robot. We used two microphone configurations: four quasi-planar microphones, located on the top of the head, numbered 5, 8, 11, 12, and eight microphones numbered 1, 3, 4, 5, 8, 10, 11, 12. An optical motion capture system was used to provide ground-truth positions of the robot and  of the speakers. The participants speak continuously during the entire recordings. However, speech pauses are inevitable and these pauses may last several seconds. Each participant has a head-mounted microphone. We applied the voice activity detector \cite{li2016iwaenc} to these microphone signals to obtain ground-truth voice activity information of each participant.
\addnote[snr1]{1}{The \ac{SNR} is approximatively 23.4~dB}

The \ac{Kinovis-MST} dataset was recorded in the Kinovis multiple-camera laboratory at INRIA Grenoble.\footnote{The Kinovis-MST dataset is publicly available at: \url{https://team.inria.fr/perception/the-kinovis-mst-dataset/}} The room size is $10.19$~m $\times$ $9.87$~m $\times$ $5.6$~m, with $T_{60}\approx0.53$~s. A v5 NAO robot with four microphones \cite{li2016iros} was used. The geometric layout of the microphones is similar to the one of the robot used in LOCATA. The speakers were moving around the robot with a speaker-to-robot distance ranging between $1.5$~m and $3.5$~m. As with LOCATA, a motion capture system was  used to obtain ground-truth trajectories of the moving participants and the location of the robot. Ten sequences were recorded with up to three participants, for a total length of about 357~s. \addnote[snr2]{1}{The robot's head has built-in fans located nearby the microphones, hence the recordings contain a notable amount of stationary and spatially correlated noise with an \ac{SNR} of approximatively 2.7~dB\cite{li2016iros}}. The participants behave more naturally than in the LOCATA scenarios, i.e. they take speech turns in a natural multi-party dialog. When one participant is silent, he/she manually hides the infrared marker located on his head to make it invisible to the motion capture system. This provides ground-truth speech activity information for each participant. This dataset and the associated annotations allow us to test the proposed tracking algorithm when the number of active speakers varies over time.

\begin{figure}[t]
\centering
{\includegraphics[width=0.7\columnwidth]{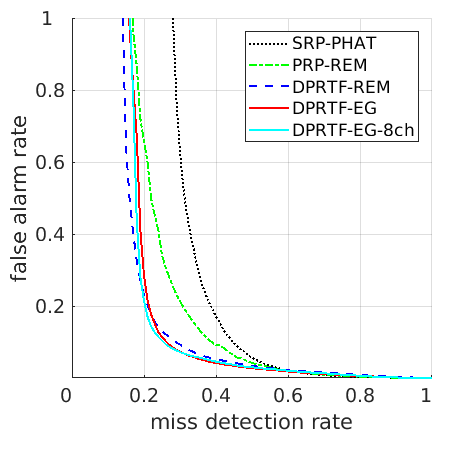}} 
\caption{\small{ROC curve for the LOCATA dataset.}} 
\label{fig:locata-roc}
\end{figure}

\subsubsection{Parameter setting}

For both datasets, we perform $360^\circ$-wide azimuth estimation and tracking: $D=72$ azimuth directions at every $5^\circ$ in [$-175^\circ$, $180^\circ$] are used as candidate directions. The \ac{CGMM} mean $c_f^{i,d}$ is the \ac{HRTF} ratio between two microphones, which are precomputed based on the direct-path propagation model for each candidate direction. In the \ac{Kinovis-MST} dataset, the \acp{HRTF} have been measured to  compute the \ac{CGMM} means. For LOCATA, the \acp{TDOA} are  computed based on the coordinate of microphones, which are then   used to compute the phase of the \ac{CGMM} means, while the magnitude of the \ac{CGMM} means are set to a constant, e.g. 0.5, for all the frequencies. 
All the recorded signals are resampled to $16$ kHz. The \ac{STFT} uses the Hamming window with length of $16$~ms and shift of $8$~ms. The \ac{CTF} length is $Q=8$ frames. The RLS forgetting factor $\lambda$ is computed using $\rho=1$. The smoothing factor $\beta$ is set to $0.9$. The exponentiated gradient update factor is $\eta=0.07$. The smoothing factor $\eta'$ is set to 0.065.  The entropy regularization factor is $\gamma=0.1$. 
For the tracker, the covariance matrix is set to be isotropic $\Sigmamat = 0.03\Imat_{2}$. The threshold giving birth to a new identity is $\tau_{1} = 0.75$ and  $L = 3$.  To decide whether a person is speaking or is silent, $L' = 3$ frames are used, with a threshold $\delta = 0.15$. At each time instance, the \ac{VEM} algorithm has $5$ iterations. Corresponding to the \ac{STFT} frame shift, i.e. 8 ms, the frame rate of the proposed system is 125 frames per second.  

\subsubsection{Comparison with Baseline Methods}
The proposed method is evaluated both in ``frame-wise localization'' mode and in ``tracker'' mode. In the first mode, the frame-wise online localization module of Section~\ref{sec:eg} is applied without the tracker of Section~\ref{sec:tracking}. Instead, it is followed by the peak selection process described in \cite{li2017taslp}. This method is referred to as \ac{DPRTF-EG}. In tracker mode, \ac{DPRTF-EG} is directly followed by the proposed \ac{VEM} tracker, without peak selection. It is then simply referred to as \ac{VEM}-tracker. In that case, the directions of active speakers are given by the state variable, and the continuity of the speaker tracks is given by the assignment variable. 
We compare \ac{DPRTF-EG} with several baseline methods: 
\begin{itemize}[leftmargin=*]
\item
The standard beamforming-based localization method called \ac{SRP} using \ac{PHAT} (\ac{SRP}-\ac{PHAT}) \cite{dibiase2001}.  The same \ac{STFT} configuration and candidate directions  are used for \ac{SRP}-\ac{PHAT} and for the proposed method. The steering vector for each candidate direction is derived from the \acp{HRTF} and \acp{TDOA} for the \ac{Kinovis-MST} and LOCATA datasets, respectively. The frame-wise \ac{SRP} is recursively smoothed with a smoothing factor set to $0.065$. 
\item A method combining PRP features, \ac{CGMM} model and parameter update using \ac{REM} \cite{schwartz2014}, referred to as PRP-REM. We also combine the DPRTF features and \ac{CGMM} with \ac{REM} (referred to as DPRTF-REM). This is to evaluate the proposed \ac{DP-RTF} feature w.r.t. PRP, and the \ac{EG}-based online parameters update method w.r.t. \ac{REM}. For both baselines, the \ac{STFT} and \ac{CGMM} settings are the same as for the proposed method. The updating factor of \ac{REM} is set to $0.065$. 
\end{itemize}

\subsubsection{Evaluation Metrics}
The detected speakers should be assigned to the actual speakers for performance evaluation. This is done using a greedy matching algorithm. First the azimuth difference for all possible detected-actual speaker pairs are computed, then the detected-actual speaker pair with the smallest difference is picked out as a matched pair. This procedure is iterated until the detected or actual speakers are all picked out.   
For each matched pair, the detected speaker is then considered to be successfully localized if the azimuth difference is not larger than $15^{\circ}$. 
The absolute error is calculated for the successfully localized sources. The \ac{MAE} is computed by averaging the absolute error of all speakers and frames. For the unsuccessful localizations, we count the \ac{MD} (speaker active but not detected) and \acp{FA} (speaker detected but not active). Then the \ac{MD} and \ac{FA} rates are computed, using all the frames, as the percentage of the total \acp{MD} and \acp{FA} out of the total number of actual speakers, respectively. In addition to these localization metrics, we also count the \acp{ID} to evaluate the tracking continuity. \ac{ID} is an absolute number. It represents the number of the identity changes in the tracks for a whole test sequence.     

The computation time is measured with the \ac{RF}, which is the processing time of a method divided by the length of the processed signal. Note that all the methods are implemented in MATLAB.

    \begin{figure*}[t]
\centering
\subfloat[Ground truth]{\includegraphics[width=0.43\textwidth]{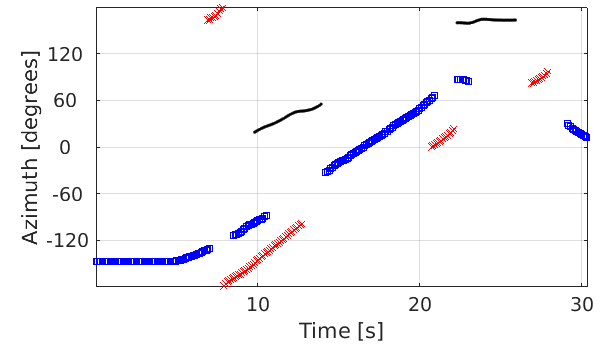}} 
\subfloat[SRP-PHAT]{\includegraphics[width=0.43\textwidth]{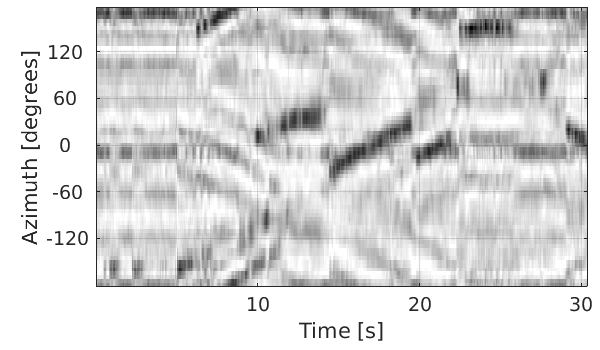}} \\
\subfloat[PRP-REM]{\includegraphics[width=0.43\textwidth]{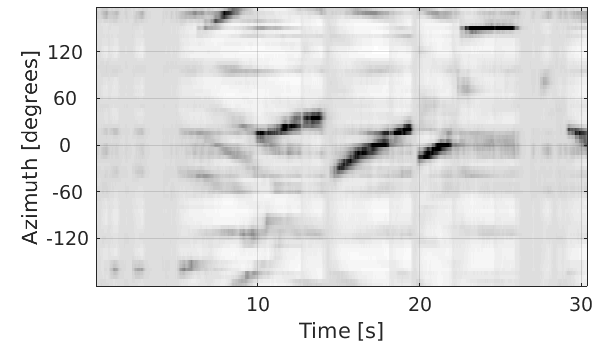}} 
\subfloat[DPRTF-REM]{\includegraphics[width=0.43\textwidth]{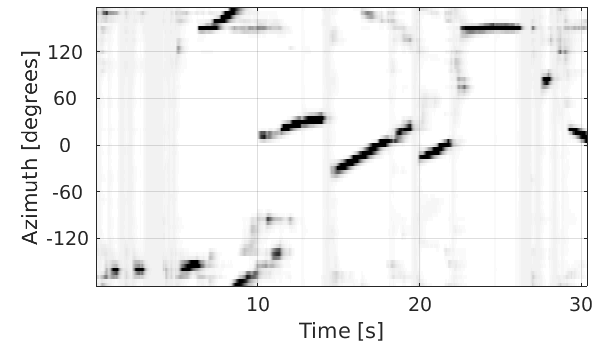}} \\
\subfloat[DPRTF-EG]{\includegraphics[width=0.43\textwidth]{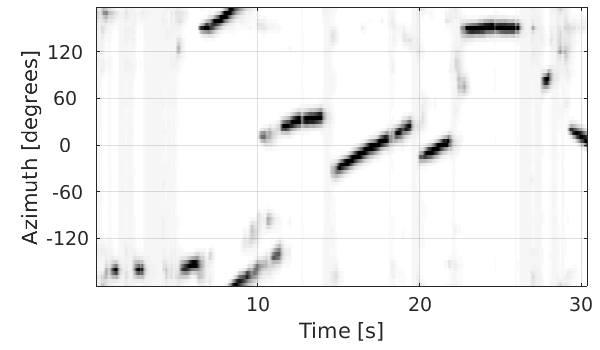}} 
\subfloat[VEM-tracking]{\includegraphics[width=0.43\textwidth]{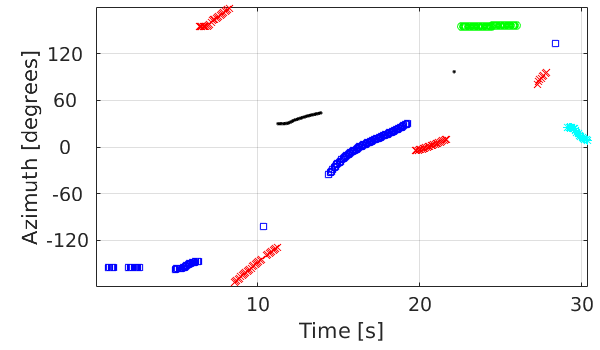}} \\
\caption{\small{Results of speaker localization and tracking for one sequence of the \ac{Kinovis-MST} dataset. (a) Ground truth trajectory and voice activity (red for speaker 1, black for speaker 2, blue for speaker 3). (b)-(e) One-dimensional heat maps as a function of time for the four tested localization methods. (f) Results for the proposed \ac{VEM}-based tracker.}}
\label{fig:kinovis-example}
\end{figure*}

\subsection{Results for LOCATA Dataset}

For convenience, both the spatial spectrum of \ac{SRP}-\ac{PHAT} and the \ac{CGMM} component weights profile will be referred to as heatmaps. Fig.~\ref{fig:locata_example} shows an example of a result obtained with a LOCATA sequence. 
Two speakers are moving and continuously speaking with short pauses. The \ac{SRP}-\ac{PHAT} heatmap (Fig.~\ref{fig:locata_example}~(b)) is cluttered due to the non ideal beampattern of the microphone array and to the influence of reverberation and noise. For most of the time, \ac{SRP}-\ac{PHAT} has prominent response power for the true speaker directions. Localization of the most dominant speaker can be made by selecting the direction with the largest response power. However, it is difficult to correctly count the number of active speakers and localize less dominant speakers, since there exist a number of spurious peaks. PRP-REM (Fig.~\ref{fig:locata_example}~(c)) exhibits a clearer heatmap compared to \ac{SRP}-\ac{PHAT}, but there exist some spurious trajectories as well, since the PRP features are contaminated by reverberation. DPRTF-REM (Fig.~\ref{fig:locata_example}~(d)) removes most of the spurious trajectories, which illustrates the robustness of the proposed \ac{DP-RTF} feature against reverberation. From Fig.~\ref{fig:locata_example}~(e), it can be seen that the proposed \ac{EG} algorithm further removes  the interferences by applying the entropy regularization. In addition, the peak evolution is smoother compared with Fig.~\ref{fig:locata_example}~(d), which is mainly due to the use of the spatial smoothing.
 Fig.~\ref{fig:locata_example}~(f) illustrates the result obtained with the proposed \ac{VEM} tracker, with \ac{DPRTF-EG} providing the observations. The proposed tracker gives smoother and cleaner results compared with the other methods. Even when the observations have a low weight, the tracker is still able to give the correct speaker trajectories. This is ensured by the second term in \eqref{eq:posterior-distribution-of-Xtn-mean} which exploits the source dynamics model and continues to provide localization information even when $w_{t,d}$ (and/or $\alpha_{tdn}$) becomes small. 
As a result, the tracker is able to preserve the identity of speakers in spite of the (short) speech pauses. In the presented sequence example, the estimated speaker identities are quite consistent with the ground truth. 
         
To empirically evaluate the quality of the heatmaps provided by the localization methods, we computed the \ac{ROC} curve (\ac{MD} rate versus \ac{FA} rate) for the LOCATA dataset by varying the peak selection threshold, for each tested method, Fig.~\ref{fig:locata-roc}.
For the \ac{ROC} curve, the closer to the left-bottom the better. 
\addnote[ch8]{1}{As already mentioned, in addition to using four microphones, we also tested an eight-microphone configuration, which is referred to as  \ac{DPRTF-EG}-8ch}. 

By analyzing the \ac{ROC} curves, one notices that the methods based on \ac{DP-RTF} perform better than \ac{SRP}-\ac{PHAT} and than PRP-REM, which is consistent with the heatmaps of Fig.~\ref{fig:locata_example}: \ac{SRP}-\ac{PHAT} and PRP-REM are more sensitive to the presence of reverberations than the proposed methods. The performance of both DPRTF-REM and \ac{DPRTF-EG} cannot be easily discriminated using the \ac{ROC} curves. \addnote[ch81]{1}{\ac{DPRTF-EG}-8ch performs slightly better than \ac{DPRTF-EG}, which means that the performance of the proposed method can be slightly improved by increasing the number of microphones. One may conclude that the proposed method is well suited when only a small number of microphones are available}.
 \addnote[roc]{1} {With all methods, the \ac{FA} rate can be trivially decreased to be close to 0 by increasing the peak selection threshold. However, the \ac{MD} rate cannot be decreased to 0 even with a very small peak-selection threshold, since some speech frames that are actually present cannot be detected as the heatmap peaks due to the influence of noise and reverberation, and to a possible latency in the detection.} 

\begin{figure}[t]
\centering
{\includegraphics[width=0.75\columnwidth]{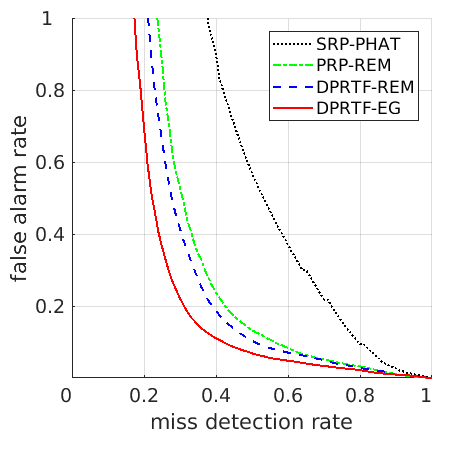}} 
\caption{\small{ROC curve for the \ac{Kinovis-MST} dataset.}} 
\label{fig:kinovis-roc}
\end{figure} 

\setlength{\tabcolsep}{4.0pt}
\begin{table}[h!]
\centering
\caption{\small Localization and tracking results for the LOCATA data.}
\label{tab:locata}
\begin{tabular}{c | c  c   c c    c  }   \vspace{1mm}
	       & MD rate (\%)  & FA rate (\%)  & MAE ($^\circ$)   & IDs & RF \\  
\ac{SRP}-\ac{PHAT}   &  39.2     & 18.6      & 5.2              & -  & 0.06 \\  
PRP-REM    &  30.9     & 19.6      & 5.0              & -  & 0.30 \\ 
DPRTF-REM  &  23.3	   & 15.2      & 4.6              & -  & 0.97 \\ 
\ac{DPRTF-EG}   &  23.9     & 13.0      & 4.0              & -  & 0.97 \\
\ac{DPRTF-EG}-8ch & 22.7    & 13.2      & 4.1              & -  & 3.03 \\ 
\ac{VEM} + EG   &  22.7     & 12.4 	   & 4.1    &  10 & 2.05\\
\ac{VEM} + EG-8ch     &  22.9     & 11.0 	   & 3.2    &  6 & 4.05\\
\end{tabular}
\end{table} 

For each curve, a good balance between \ac{FA} rate and \ac{MD} rate is achieved at the left-bottom corner, which can be detected as the point with the minimum distance to the origin. The average localization results corresponding to this optimal left-bottom point are summarized in Table~\ref{tab:locata} for each tested method. It can be seen that, besides \ac{MD} and \ac{FA}, the DPRTF-based methods achieve smaller \ac{MAE} than \ac{SRP}-\ac{PHAT} and PRP-REM, since the proposed \ac{DP-RTF} features are robust against reverberation and thus leads to smaller biases for the heatmap peaks. \ac{DPRTF-EG} has a higher \ac{MD} rate than DPRTF-REM, while it also has lower \ac{FA} rate, and a lower \ac{MAE}, due to the effect of entropy regularization. 
\addnote[ch82]{1}{With eight microphones, i.e. \ac{DPRTF-EG}-8ch, \ac{MD} is 1\% smaller than the \ac{MD} of \ac{DPRTF-EG}, since the use of a non coplanar microphone setup provides more accurate localization than a coplanar setup}. 
The proposed tracker performs the best in terms of \ac{MD} and of \ac{FA}. For the four-microphone configuration, the tracker slightly reduces \ac{FA} compared to \ac{DPRTF-EG}. It also reduces the \ac{MD} score since some correct speaker trajectories can be recovered even when the observations have (very) low weights, as explained above. \addnote[np1]{2}{In addition,  the \ac{MAE} is noticeably reduced when more microphones are used by the \ac{VEM} tracker, which is not the case with the DPRTF-EG localizer. This phenomenon indicates that, compared with the localizer, the tracker is able to better exploit additional information available with extra microphones, namely to revise the speaker trajectory estimation, since the state dynamics of the tracker helps correcting the possibly inaccurate additional localization information}. 
The proposed tracker achieves quite consistent speaker \ac{ID} estimation. \addnote[ch83]{1}{For the whole LOCATA dataset, only ten identity switches were observed when using \ac{DPRTF-EG}, and this number is reduced to six when using \ac{DPRTF-EG}-8ch}.  The remaining identity switches are mainly due to speakers with crossing trajectories, a hard case for multiple audio-source tracking.

As for the computation time, \ac{SRP}-\ac{PHAT} has the smallest \ac{RF}. Based on the fact that the \acp{RF} of DPRTF-REM and \ac{DPRTF-EG} are identical, we can conclude that the \ac{REM} algorithm and the proposed \ac{EG} algorithm have comparable computational complexities. The \acp{RF} of PRP-REM, DPRTF-REM (or \ac{DPRTF-EG}) and \ac{DPRTF-EG}-8ch are different due to different computational complexities for feature estimation, more precisely due to the different dimensions of the vector to be estimated.  The \ac{CTF} identification used for \ac{DP-RTF} estimation solves an \ac{RLS} problem with the unknown \ac{CTF} vector $\tilde{\mathbf{a}}_f\in \mathbb{C}^{(IQ-1)\times1}$. Remind that $I$ and $Q$ denote the number of microphones and the \ac{CTF} length, respectively. In the present work, we have set $I=4$/$Q=8$ for DPRTF-REM (or \ac{DPRTF-EG}), $I=8$/$Q=8$ for \ac{DPRTF-EG}-8ch. \ac{PRP} is defined based on the narrow-band assumption, or equivalently based on the \ac{CTF} with $Q=1$, thence we have $I=4$/$Q=1$ for PRP-REM. The proposed localization method, i.e. \ac{DPRTF-EG} with four microphones, has an \ac{RF} smaller than one, which means it can be run in real time. The \ac{RF} for the proposed tracker (\ac{VEM}) is computed by the sum of the localization time and of the tracking time. For acoustic tracking, the tracker observes an \ac{DOA} estimate every 8~ms. However, an 8~ms speaker motion is small. Thus in practice, the tracker uses one \ac{DOA} estimate per 32~ms  intervals, which leads to an \ac{RF} of 2.05 for the \ac{4ch} case and 4.05 for the \ac{8ch} case. The \ac{RF} can be further improved by using less \ac{DOA} estimates.

\subsection{Results for \ac{Kinovis-MST} Dataset} 
\setlength{\tabcolsep}{4.0pt}
\begin{table}[t]
\centering
\caption{\small Localization and tracking results for the \ac{Kinovis-MST} dataset.}
\label{tab:kinovis}
\begin{tabular}{c | c  c   c c  c    }   \vspace{1mm}
	       & MD rate (\%)  & FA rate (\%)  & MAE ($^\circ$)   & IDs & \ac{RF}  \\  
\ac{SRP}-\ac{PHAT}   &  60.0     & 37.1      & 5.5              & - & 0.07  \\  
PRP-REM    &  40.3     & 23.1      & 5.1              & - & 0.32 \\ 
DPRTF-REM  &  37.6	   & 22.0      & 5.5              & - & 0.73 \\ 
\ac{DPRTF-EG}   &  31.4     & 19.5      & 5.3              & - & 0.73 \\
\ac{VEM} + EG  &  31.1     & 11.7      &  4.9             & 11 & 2.12
\end{tabular}
\end{table}   


Fig.~\ref{fig:kinovis-example}  shows an example of result for a \ac{Kinovis-MST} sequence. 
Three participants are moving and intermittently speaking. It can be seen that, for many frames, the response power of \ac{SRP}-\ac{PHAT} and the \ac{CGMM} component weights of PRP-REM corresponding to the true active speakers are not prominent, compared to the spurious trajectories. Again, DPRTF-REM and \ac{DPRTF-EG} provide much better heatmaps, though they also miss some speaking frames, e.g. at the beginning of Speaker 3's trajectory (in blue). The possible reasons are i) the NAO robot (v5) has a relative strong ego-noise \cite{li2016iros}, and thus the signal-to-noise ratio of the recorded signals is relative low, and ii) the speakers are moving with a varying source-to-robot distance and the direct-path speech is contaminated by more reverberations when the speakers are distant. Overall, DPRTF-REM and \ac{DPRTF-EG} are able to monitor the moving, appearance, and disappearance of active speakers for most of the time, with a small time lag due to the temporal smoothing.


This kind of recording/scenario is very challenging for the tracking method, especially for speaker identity preservation, since the participants are intermittently speaking and moving. In a general manner, the proposed tracker achieves relatively good results, as illustrated in Fig.~\ref{fig:kinovis-example} (f). The tracked trajectories are smooth and clean. If the true trajectory of one speaker has an approximately constant direction, the tracker is able to re-identify the speaker even after a long silence thanks to the above-mentioned combination of observations and dynamics in \eqref{eq:posterior-distribution-of-Xtn-mean}, e.g. Speaker 1's trajectory in red. In the case that the speaker changes his/her movement when he/she is silent, the track can be lost. When the person speaks again, it is indeed difficult to re-idendify him/her based on the dynamics estimated before the silence period. The tracker may then prefer to give birth to a new speaker. This is illustrated by the black trajectory turning into green, and the blue trajectory turning into cyan in Fig.~\ref{fig:kinovis-example}. Note that the silence periods are here much longer than in the LOCATA example of Fig.~\ref{fig:locata_example}. 

Fig~\ref{fig:kinovis-roc} show the \ac{ROC} curves for the \ac{Kinovis-MST} dataset. Compared to the \ac{ROC} curves for the LOCATA dataset,  all the four localization methods have a worse \ac{ROC} curve, especially along the \ac{MD} rate axis, for the reasons mentioned above.
Table \ref{tab:kinovis} summarizes the localization and tracking results for the optimal bottom-left point of the \ac{ROC} curves. It can be seen that, for the four localization methods, \acp{MAE} are quite close, namely the heatmap peaks have similar biases.
Compared with the results for the LOCATA dataset, the advantage of the proposed tracker is more significant for the \ac{Kinovis-MST} dataset. \addnote[np2]{2}{In particular, the \ac{FA} rate is reduced by 7.8\% relatively to \ac{DPRTF-EG}, and is similar to the \ac{FA} rate obtained with the LOCATA dataset. This means that the dynamic model associated with the tracker can efficiently reduce the influence of incorrect source localizations caused by noise and by complex source movements}. 
The identity switches are mainly caused by speakers changing their direction of movement while during silent periods, as discussed above.  
Compared to the LOCATA dataset, \ac{DPRTF-EG} has smaller \ac{RF}, since the \ac{Kinovis-MST} dataset is noisier and more noise frames are skipped in the RLS algorithm.

\subsection{Dicussion}
\addnote[vd]{2}{
The experimental results obtained with the two datasets clearly show the effectiveness of the proposed method based on DP-RTF estimation, multiple speaker localization and variational tracking. To improve robustness, temporal smoothing is used, which leads to localization/tracking latency. This latency causes MD and FA observed at both the beginning and the end of continuous speech segments. However, it can be observed from the examples shown in Fig.~\ref{fig:locata_example} and Fig.~\ref{fig:kinovis-example} that the latency is not that severe. The Kinovis-MST dataset is more challenging than the LOCATA dataset for speaker localization/tracking, due to its lower SNR and the presence of casual speaking style. Even though, the proposed methods achieve a comparable FA rate with the two datasets. 
Concerning the \ac{MD} score, when applied to Kinovis-MST, the method yields larger MD rates than when applied to LOCATA. This is due to the large number of TF bins dominated by a high SNR score, present in the Kinovis-MST recordings. The tracker's dynamics attenuate the influence of these TF bins to a limited extend.}


\section{Conclusion}\label{sec:conclusion}
In this paper, we proposed and combined i) a recursive \ac{DP-RTF} feature estimation method, ii) an online multiple-speaker localization method, and iii) an multiple-speaker tracking method. The resulting framework provides online speaker counting, localization and consistent tracking (\ie preserving speaker identity over a track in spite of intermittent speech production). 
The three algorithms are computationally efficient. In particular the tracking algorithm implemented in variational Bayesian framework yields a tractable solver under the form of \ac{VEM}. Experiments with two datasets, recorded in realistic environment, verify that the proposed method is robust against reverberation and noise. Moreover, the tracker is able to efficiently track multiple moving speakers, detect whether they are speech or they are silent,  as long as the motion associated with silent people is smooth. However, the tracking of the person from silent to active remains a difficult task. The combination of the proposed method with speaker identification will be addressed in the future.

The proposed \ac{VEM} tracker can be easily adapted to work in tandem with any frame-wise localizer providing source location estimates and/or corresponding weights (and if no weights are provided by the localizer, the tracker can be applied with all weights set to one). This makes the proposed tracker very flexible, and easily reusable by the audio processing research community.

\bibliographystyle{ieeetr}
\balance

\begin{thebibliography}{10}

\bibitem{knapp1976}
C.~Knapp and G.~C. Carter, ``The generalized correlation method for estimation
  of time delay,'' {\em IEEE Transactions on Acoustics, Speech and Signal
  Processing}, vol.~24, no.~4, pp.~320--327, 1976.

\bibitem{chen2006}
J.~Chen, J.~Benesty, and Y.~Huang, ``Time delay estimation in room acoustic
  environments: an overview,'' {\em EURASIP Journal on applied signal
  processing}, vol.~2006, pp.~170--170, 2006.

\bibitem{dibiase2001}
J.~H. DiBiase, H.~F. Silverman, and M.~S. Brandstein, ``Robust localization in
  reverberant rooms,'' in {\em Microphone Arrays} (M.~S. Brandstein and
  D.~Ward, eds.), pp.~157--180, Springer, 2001.

\bibitem{ishi2009}
C.~T. Ishi, O.~Chatot, H.~Ishiguro, and N.~Hagita, ``Evaluation of a
  music-based real-time sound localization of multiple sound sources in real
  noisy environments,'' in {\em IEEE/RSJ International Conference on
  Intelligent Robots and Systems (IROS)}, pp.~2027--2032, 2009.

\bibitem{yilmaz2004}
O.~Yilmaz and S.~Rickard, ``Blind separation of speech mixtures via
  time-frequency masking,'' {\em IEEE Transactions on Signal Processing,},
  vol.~52, no.~7, pp.~1830--1847, 2004.

\bibitem{mandel2010}
M.~I. Mandel, R.~J. Weiss, and D.~P. Ellis, ``Model-based
  expectation-maximization source separation and localization,'' {\em IEEE
  Transactions on Audio, Speech, and Language Processing}, vol.~18, no.~2,
  pp.~382--394, 2010.

\bibitem{dorfan2015}
Y.~Dorfan and S.~Gannot, ``Tree-based recursive expectation-maximization
  algorithm for localization of acoustic sources,'' {\em IEEE/ACM Transactions
  on Audio, Speech, and Language Processing}, vol.~23, no.~10, pp.~1692--1703,
  2015.

\bibitem{huang2003adaptive}
Y.~Huang and J.~Benesty, ``Adaptive multichannel time delay estimation based on
  blind system identification for acoustic source localization,'' in {\em
  Adaptive Signal Processing}, pp.~227--247, Springer, 2003.

\bibitem{doclo2003}
S.~Doclo and M.~Moonen, ``Robust adaptive time delay estimation for speaker
  localization in noisy and reverberant acoustic environments,'' {\em EURASIP
  Journal on Applied Signal Processing}, vol.~2003, pp.~1110--1124, 2003.

\bibitem{dvorkind2005}
T.~G. Dvorkind and S.~Gannot, ``Time difference of arrival estimation of speech
  source in a noisy and reverberant environment,'' {\em Signal Processing},
  vol.~85, no.~1, pp.~177--204, 2005.

\bibitem{kowalczyk2013}
K.~Kowalczyk, E.~A. Habets, W.~Kellermann, and P.~A. Naylor, ``Blind system
  identification using sparse learning for {TDOA} estimation of room
  reflections,'' {\em IEEE Signal Processing Letters}, vol.~20, no.~7,
  pp.~653--656, 2013.

\bibitem{li2017taslp}
X.~Li, L.~Girin, R.~Horaud, and S.~Gannot, ``Multiple-speaker localization
  based on direct-path features and likelihood maximization with spatial
  sparsity regularization,'' {\em IEEE/ACM Transactions on Audio, Speech, and
  Language Processing}, vol.~25, no.~10, pp.~1997--2012, 2017.

\bibitem{avargel2007}
Y.~Avargel and I.~Cohen, ``System identification in the short-time {F}ourier
  transform domain with crossband filtering,'' {\em IEEE Transactions on Audio,
  Speech, and Language Processing}, vol.~15, no.~4, pp.~1305--1319, 2007.

\bibitem{talmon2009}
R.~Talmon, I.~Cohen, and S.~Gannot, ``Relative transfer function identification
  using convolutive transfer function approximation,'' {\em IEEE Transactions
  on Audio, Speech, and Language Processing}, vol.~17, no.~4, pp.~546--555,
  2009.

\bibitem{pavlidi2013}
D.~Pavlidi, A.~Griffin, M.~Puigt, and A.~Mouchtaris, ``Real-time multiple sound
  source localization and counting using a circular microphone array,'' {\em
  IEEE Transactions on Audio, Speech, and Language Processing}, vol.~21,
  no.~10, pp.~2193--2206, 2013.

\bibitem{schwartz2014}
O.~Schwartz and S.~Gannot, ``Speaker tracking using recursive {EM}
  algorithms,'' {\em IEEE/ACM Transactions on Audio, Speech, and Language
  Processing}, vol.~22, no.~2, pp.~392--402, 2014.

\bibitem{roman2008}
N.~Roman and D.~Wang, ``Binaural tracking of multiple moving sources,'' {\em
  IEEE Transactions on Audio, Speech, and Language Processing}, vol.~16, no.~4,
  pp.~728--739, 2008.

\bibitem{evers2015}
C.~Evers, A.~H. Moore, P.~A. Naylor, J.~Sheaffer, and B.~Rafaely,
  ``Bearing-only acoustic tracking of moving speakers for robot audition,'' in
  {\em IEEE International Conference on Digital Signal Processing (DSP)},
  pp.~1206--1210, 2015.

\bibitem{ban2017}
Y.~Ban, L.~Girin, X.~Alameda-Pineda, and R.~Horaud, ``Exploiting the
  complementarity of audio and visual data in multi-speaker tracking,'' in {\em
  ICCV Workshop on Computer Vision for Audio-Visual Media}, vol.~3, 2017.

\bibitem{liang2008robust}
Z.~Liang, X.~Ma, and X.~Dai, ``Robust tracking of moving sound source using
  multiple model kalman filter,'' {\em Applied acoustics}, vol.~69, no.~12,
  pp.~1350--1355, 2008.

\bibitem{vermaak2001nonlinear}
J.~Vermaak and A.~Blake, ``Nonlinear filtering for speaker tracking in noisy
  and reverberant environments,'' in {\em Acoustics, Speech, and Signal
  Processing, 2001. Proceedings.(ICASSP'01). 2001 IEEE International Conference
  on}, vol.~5, pp.~3021--3024, IEEE, 2001.

\bibitem{ba2016line}
S.~Ba, X.~Alameda-Pineda, A.~Xompero, and R.~Horaud, ``An on-line variational
  bayesian model for multi-person tracking from cluttered scenes,'' {\em
  Computer Vision and Image Understanding}, vol.~153, pp.~64--76, 2016.

\bibitem{gebru2017audio}
I.~Gebru, S.~Ba, X.~Li, and R.~Horaud, ``Audio-visual speaker diarization based
  on spatiotemporal {Bayesian} fusion,'' {\em {IEEE Transactions on Pattern
  Analysis and Machine Intelligence}}, 2017.

\bibitem{fallon2012acoustic}
M.~F. Fallon and S.~J. Godsill, ``Acoustic source localization and tracking of
  a time-varying number of speakers,'' {\em IEEE Transactions on Audio, Speech,
  and Language Processing}, vol.~20, no.~4, pp.~1409--1415, 2012.

\bibitem{valin2007robust}
J.-M. Valin, F.~Michaud, and J.~Rouat, ``Robust localization and tracking of
  simultaneous moving sound sources using beamforming and particle filtering,''
  {\em Robotics and Autonomous Systems}, vol.~55, no.~3, pp.~216--228, 2007.

\bibitem{cevher2007acoustic}
V.~Cevher, R.~Velmurugan, and J.~H. McClellan, ``Acoustic multitarget tracking
  using direction-of-arrival batches,'' {\em IEEE Transactions on Signal
  Processing}, vol.~55, no.~6, pp.~2810--2825, 2007.

\bibitem{vo2004tracking}
B.-N. Vo, S.~Singh, and W.~K. Ma, ``Tracking multiple speakers using random
  sets,'' in {\em Acoustics, Speech, and Signal Processing, 2004.
  Proceedings.(ICASSP'04). IEEE International Conference on}, vol.~2,
  pp.~ii--357, IEEE, 2004.

\bibitem{ma2006tracking}
W.-K. Ma, B.-N. Vo, S.~S. Singh, and A.~Baddeley, ``Tracking an unknown
  time-varying number of speakers using tdoa measurements: A random finite set
  approach,'' {\em IEEE Transactions on Signal Processing}, vol.~54, no.~9,
  pp.~3291--3304, 2006.

\bibitem{vo2006}
B.-N. Vo and W.-K. Ma, ``The gaussian mixture probability hypothesis density
  filter,'' {\em IEEE Transactions on signal processing}, vol.~54, no.~11,
  pp.~4091--4104, 2006.

\bibitem{li2018sam}
X.~Li, B.~Mourgue, L.~Girin, S.~Gannot, and R.~Horaud, ``Online localization of
  multiple moving speakers in reverberant environments,'' in {\em The Tenth
  IEEE Workshop on Sensor Array and Multichannel Signal Processing}, 2018.

\bibitem{kivinen1997}
J.~Kivinen and M.~K. Warmuth, ``Exponentiated gradient versus gradient descent
  for linear predictors,'' {\em Information and Computation}, vol.~132, no.~1,
  pp.~1--63, 1997.

\bibitem{xu1995}
G.~Xu, H.~Liu, L.~Tong, and T.~Kailath, ``A least-squares approach to blind
  channel identification,'' {\em IEEE Transactions on signal processing},
  vol.~43, no.~12, pp.~2982--2993, 1995.

\bibitem{li2015icassp}
X.~Li, L.~Girin, R.~Horaud, and S.~Gannot, ``Estimation of relative transfer
  function in the presence of stationary noise based on segmental power
  spectral density matrix subtraction,'' in {\em IEEE International Conference
  on Acoustics, Speech and Signal Processing}, pp.~320--324, 2015.

\bibitem{yuille2003}
A.~L. Yuille and A.~Rangarajan, ``The concave-convex procedure,'' {\em Neural
  computation}, vol.~15, no.~4, pp.~915--936, 2003.

\bibitem{gebru2016algorithms}
I.~D. Gebru, X.~Alameda-Pineda, F.~Forbes, and R.~Horaud, ``Em algorithms for
  weighted-data clustering with application to audio-visual scene analysis,''
  {\em IEEE transactions on pattern analysis and machine intelligence},
  vol.~38, no.~12, pp.~2402--2415, 2016.

\bibitem{LOCATA2018a}
H.~W. L{\"o}llmann, C.~Evers, A.~Schmidt, H.~Mellmann, H.~Barfuss, P.~A.
  Naylor, and W.~Kellermann, ``The {LOCATA} challenge data corpus for acoustic
  source localization and tracking,'' in {\em {IEEE Sensor Array and
  Multichannel Signal Processing Workshop}}, (Sheffield, UK), July 2018.

\bibitem{li2016iwaenc}
X.~Li, R.~Horaud, L.~Girin, and S.~Gannot, ``Voice activity detection based on
  statistical likelihood ratio with adaptive thresholding,'' in {\em IEEE
  International Workshop on Acoustic Signal Enhancement (IWAENC)}, pp.~1--5,
  2016.

\bibitem{li2016iros}
X.~Li, L.~Girin, F.~Badeig, and R.~Horaud, ``Reverberant sound localization
  with a robot head based on direct-path relative transfer function,'' in {\em
  IEEE/RSJ International Conference on Intelligent Robots and Systems (IROS)},
  pp.~2819--2826, IEEE, 2016.

\end{thebibliography}

\end{document}